\documentclass[]{aa}
\usepackage[varg]{txfonts}
\bibpunct{(}{)}{;}{a}{}{,} 
\begin{document}
\title{Massive pulsating stars observed by BRITE-Constellation\thanks{Based on data collected by the BRITE-Constellation satellite mission, built, launched and operated thanks to support from the Austrian Aeronautics and Space Agency and the University of Vienna, the Canadian Space Agency (CSA) and the Foundation for Polish Science \& Technology (FNiTP MNiSW) and National Centre for Science (NCN).}}
\subtitle{I. The triple system $\beta$~Centauri (Agena)}
\titlerunning{$\beta$~Centauri}
\author{A.~Pigulski\inst{\ref{inst1}}\and H.~Cugier\inst{\ref{inst1}}\and A.~Popowicz\inst{\ref{inst2}}\and R.~Kuschnig\inst{\ref{inst3}}\and A.~F.~J.~Moffat\inst{\ref{inst4}}\and S.~M.~Rucinski\inst{\ref{inst5}}\and A.~Schwarzenberg-Czerny\inst{\ref{inst6}}\and W.~W.~Weiss\inst{\ref{inst3}}\and G.~Handler\inst{\ref{inst6}}\and G.~A.~Wade\inst{\ref{inst7}}\and O.~Koudelka\inst{\ref{inst8}}\and J.~M.~Matthews\inst{\ref{inst9}}\and St.~Mochnacki\inst{\ref{inst5}}\and P.~Orlea\'nski\inst{\ref{inst10}}\and H.~Pablo\inst{\ref{inst4}}\and T.~Ramiaramanantsoa\inst{\ref{inst4}}\and G.~Whittaker\inst{\ref{inst5}}\and E.~Zoc{\l}o\'nska\inst{\ref{inst6}}\and K.~Zwintz\inst{\ref{inst11}}}
\institute{Instytut Astronomiczny, Uniwersytet Wroc{\l}awski, Kopernika 11, 51-622 Wroc{\l}aw, Poland\\\email{pigulski@astro.uni.wroc.pl}\label{inst1}
\and Instytut Automatyki, Politechnika \'Sl\k{a}ska, Akademicka 16, 44-100 Gliwice, Poland\label{inst2}
\and Institut f\"ur Astrophysik, Universit\"at Wien, T\"urkenschanzstrasse 17, 1180 Wien, Austria\label{inst3}
\and D\'epartement de physique, Universit\'e de Montr\'eal, C.P.~6128, Succursale Centre-Ville, Montr\'eal, Qu\'ebec, H3C\,3J7, Canada, et Centre de recherche en astrophysique du Qu\'ebec (CRAQ)\label{inst4}
\and Department of Astronomy \& Astrophysics, University of Toronto, 50 St.~George Street, Toronto, Ontario, M5S\,3H4, Canada\label{inst5}
\and Centrum Astronomiczne im.~M.\,Kopernika, Polska Akademia Nauk, Bartycka 18, 00-716 Warszawa, Poland\label{inst6}
\and Department of Physics, Royal Military College of Canada, PO Box 17000, Station Forces, Kingston, Ontario, K7K\,7B4, Canada\label{inst7}
\and Institut f\"ur Kommunikationsnetze und Satellitenkommunikation, Technische Universit\"at Graz, Inffeldgasse 12, 8010 Graz, Austria\label{inst8}
\and Dept.~of Physics \& Astronomy, The University of British Columbia, 6224 Agricultural Road, Vancouver, B.C., V6T\,1Z1, Canada\label{inst9}
\and Centrum Bada\'n Kosmicznych, Polska Akademia Nauk, Bartycka 18A, 00-716 Warszawa, Poland\label{inst10}
\and Institut f\"ur Astro- und Teilchenphysik, Universit\"at Innsbruck, Technikerstrasse 25/8, 6020 Innsbruck, Austria\label{inst11}
}
\date{Received date / Accepted date }
\abstract {Asteroseismology of massive pulsating stars of $\beta$~Cep and SPB types can help us to uncover the internal structure of massive stars and understand certain physical phenomena that are taking place in their interiors. We study $\beta$~Centauri (Agena), a triple system with two massive fast-rotating early B-type components which show $p$- and $g$-mode pulsations; the system's secondary is also known to have a measurable magnetic field.}
{This paper aims to precisely determine the masses and detect pulsation modes in the two massive components of $\beta$~Cen with BRITE-Constellation photometry. In addition, seismic models for the components are considered and the effects of fast rotation are discussed. This is done to test the limitations of seismic modeling for this very difficult case.}
{A simultaneous fit of visual and spectroscopic orbits is used to self-consistently derive the orbital parameters, and subsequently the masses, of the components. Time-series analysis of BRITE-Constellation data is used to detect pulsation modes and derive their frequencies, amplitudes, phases, and rates of frequency change. Theoretically-predicted frequencies are calculated for the appropriate evolutionary models and their stability is checked. The effects of rotational splitting and coupling are also presented.}
{The derived masses of the two massive components are equal to 12.02\,$\pm$\,0.13 and 10.58\,$\pm$\,0.18~M$_\odot$. The parameters of the wider, A\,--\,B system, presently approaching periastron passage, are constrained. Analysis of the combined blue- and red-filter BRITE-Constellation photometric data of the system revealed the presence of 19 periodic terms, of which eight are likely $g$ modes, nine are $p$ modes, and the remaining two are combination terms. It cannot be excluded that one or two low-frequency terms are rotational frequencies. It is possible that both components of $\beta$~Cen are $\beta$~Cep/SPB hybrids. An attempt to use the apparent changes of frequency to distinguish which modes originate in which component did not succeed, but there is potential for using this method when more BRITE data become available.} 
{Agena seems to be one of very few rapidly rotating massive objects with rich $p$- and $g$-mode spectra, and precisely known masses. It can therefore be used to gain a better understanding of the excitation of pulsations in relatively rapidly rotating stars and their seismic modeling. Lacking proper mode identification, the pulsation frequencies found in $\beta$~Cen cannot yet be used to constrain the internal structure of the components, but it may be possible to achieve this in the future with the use of spectroscopy and spectropolarimetry. In particular, these kinds of data can be used for mode identification since they provide new radial velocities. In consequence, they may help to improve the orbital solution, derive more precise masses, magnetic field strength and geometry, inclination angles, and reveal rotation periods. They may also help to assign pulsation frequencies to components. Finally, the case studied here illustrates the potential of BRITE-Constellation data for the detection of rich-frequency spectra of small-amplitude modes in massive pulsating stars.}
\maketitle
\titlerunning{$\beta$~Centauri}
\authorrunning{A.~Pigulski et al.}
\section{Introduction}
 At the main sequence stage of evolution and shortly beyond, massive stars may become pulsationally unstable owing to the $\kappa$~mechanism which drives both pressure ($p$) and gravity ($g$) modes \citep{1993MNRAS.262..204D,1993MNRAS.265..588D,1999AcA....49..119P}. This presents an opportunity to probe their interiors by means of asteroseismology, and to decipher the internal rotation profile, the extent of core overshooting, and to test the input physics, especially the stellar opacities. This latter issue can indeed be studied well with massive pulsators because pulsations in these stars occur as a result of a subtle supremacy of driving over damping.

Observationally, the $p$- and $g$-mode massive pulsators are classified as $\beta$~Cep and slowly pulsating B-type (SPB) stars, respectively. The number of known massive pulsators suitable for seismic modeling, i.e.~having rich frequency spectra, is presently small. In the last decade, ground-based campaigns have revealed only a handful of $\beta$~Cep-type stars with a substantial (of the order of ten) number of detected modes. The results of the analysis of these observations were used in the subsequent attempts of seismic modeling, see reviews of case studies by \cite{2007CoAst.150..159P} and \cite{2011IAUS..272..433D}. Along with many interesting results, the case studies reveal that some observed modes, especially the low-degree $g$ modes, are usually stable in models calculated with presently-available stellar opacities. This indicates a possible need for revision of the opacities, see \cite{2004MNRAS.350.1022P}, \cite{2007MNRAS.375L..21M}, \cite{2008CoAst.157..385Z}, and \cite{2013MNRAS.432..822W,2015A&A...580L...9W}. The new opacity bump \citep{2012A&A...547A..42C,2014IAUS..301..401C,2014A&A...565A..76C} might help to solve the problem. In view of the discrepancies between the results of the experiment of \cite{2015Natur.517...56B} and theoretical predictions, some revisions of opacities applicable to early B-type stars can be expected. See also the discussion of \cite{2015MNRAS.450....2I}.

Asteroseismology of massive stars therefore seems promising, but for a real breakthrough to happen the sample of the studied objects needs to be enlarged. The objects need to cover a range of masses, ages, rotation rates and metallicities. Since pulsations in massive stars have small photometric amplitudes in the visual domain, the only way of detecting many modes is by lowering the detection threshold. One way to achieve this goal is through time-series photometry from space. Some $\beta$~Cep, SPB and hybrid $\beta$~Cep/SPB stars have already been observed by the WIRE, MOST, CoRoT, Coriolis (SMEI experiment) and Kepler satellites. The next mission which seems to be extremely relevant for the advancement of the study of massive pulsators is the BRITE-Constellation \citep[hereafter B-C,][]{2008CoAst.152..160H}. This mission consists of five functioning BRITE (BRite Target Explorer) nano-satellites that are equipped with 3-cm refracting telescopes \citep{2014PASP..126..573W}. The great advantage of B-C is that, unlike most satellite missions, observations are carried out in two passbands. This potentially allows mode identification using amplitude ratios in the two passbands; see, e.g., \cite{2008CoAst.152..140D}. The B-C mission is capable of obtaining photometry at a mmag level for stars brighter than $V\sim$\,5~mag. The great advantage of observing such bright stars is that accompanying simultaneous time-series high resolution, high signal-to-noise spectroscopy and/or spectropolarimetry \citep{2014sf2a.conf..505N,2016MNRAS.456....2W} can easily be obtained from the ground. Spectroscopy can be used for mode identification, the determination of stellar parameters, and radial velocities. Spectropolarimetry can provide information on the presence and geometry of any magnetic field.

The present paper is the first study of a $\beta$~Cep-type star with B-C data. The chosen star is $\beta$~Centauri, which is a non-eclipsing multiple system consisting of three B-type stars, of which two are massive pulsators. The star is a very good target for this kind of study because a combination of spectroscopic and visual orbits allows a precise determination of the masses of the components. The masses are the most important parameters that strongly constrain the seismic modeling. As shown in Sect.~\ref{br-phot}, we detect 17 intrinsic modes, both $p$ and $g$, from B-C photometry. This makes the star one of a few known $\beta$~Cep/SPB hybrid stars with rich frequency spectra and attractive characteristics for seismic modeling. 

This paper is organized as follows. First we derive orbital parameters (Sect.~\ref{multipl}), which enable a new determination of the masses of the two brightest components of the system. Section \ref{br-phot} presents the results of the analysis of the B-C photometry. In Sect.~\ref{smodel}, we present seismic modeling of the star, which is followed by a final discussion and conclusions (Sect.~\ref{discon}). Finally, the Appendix \ref{prep} shows the most important steps of the data-reduction procedure which is required to correct raw B-C photometry for instrumental effects.

\section{Multiplicity of $\beta$~Cen}\label{multipl}
Agena (\object{$\beta$~Centauri}, HD\,122451, $V=$~0.6~mag) is a visual hierarchical triple (WDS\,14038$-$6022) consisting of a non-eclipsing pair of early B-type stars in an eccentric orbit (components Aa and Ab) and a more distant $\sim$3~mag fainter companion. The bright pair (component A) is classified as B1\,II \citep{1950MNRAS.110...15G,1969MNRAS.144...31B} or B1\,III \citep{1957MNRAS.117..449D,1969ApJ...157..313H}. The variability of radial velocities of $\beta$~Cen\,A was discovered by \cite{1915LicOB...8..124W} and confirmed later by \cite{1962MNRAS.124..189B}. Using spectra obtained with the 74-inch Radcliffe Observatory reflector, \cite{1967MNRAS.136...51B} found that radial velocities of the star vary with a short period of 0.1317 or 0.1520~d superimposed on a gradual variation of the $\gamma$-velocity. The latter suggested binarity of component A, which was later confirmed both spectroscopically \citep{2002A&A...384..209A} and interferometrically \citep{1974MNRAS.167..121H,1999MNRAS.302..245R,2005MNRAS.356.1362D}. 

Since the multiplicity of $\beta$~Cen plays an important role in this study, we first update the solution for the Aa\,--\,Ab components' orbit and derive constraints for orbital parameters of the wider A\,--\,B system, too.

\begin{table*}[!ht]
\caption{Parameters of the Aa\,--\,Ab orbit of $\beta$~Cen.}
\label{aab-param}
\centering 
\begin{tabular}{cccc} 
\hline\hline\noalign{\smallskip}
Parameter & Visual orbit & Spectroscopic orbit & Combined orbits \\ 
\noalign{\smallskip}\hline\noalign{\smallskip}
Orbital period, $P_{\rm orb}$ [yr] & 0.9788$^{+0.0008}_{-0.0014}$ & 0.97719$^{+0.00005}_{-0.00005}$ & 0.97720$^{+0.00004}_{-0.00004}$ \\\noalign{\smallskip}
Orbital period, $P_{\rm orb}$ [d] & 357.49$^{+0.31}_{-0.51}$& 356.911$^{+0.018}_{-0.018}$ & 356.915$^{+0.015}_{-0.015}$\\\noalign{\smallskip}
Eccentricity, $e$ &0.8201$^{+0.0031}_{-0.0021}$ & 0.8237$^{+0.0011}_{-0.0011}$ &0.8245$^{+0.0006}_{-0.0006}$ \\\noalign{\smallskip}
Longitude of periastron, $\omega$ [$^{\rm o}$] &60.1$^{+0.5}_{-0.4}$ & 61.13$^{+0.36}_{-0.34}$ &60.87$^{+0.26}_{-0.25}$ \\\noalign{\smallskip}
Time of periastron passage, $T_0$ [Besselian epoch]&2000.1509$^{+0.0010}_{-0.0010}$ & 2000.15209$^{+0.00012}_{-0.00013}$ & 2000.15202$^{+0.00011}_{-0.00010}$ \\\noalign{\smallskip}
Inclination, $i$ [$^{\rm o}$] & 66.76$^{+0.30}_{-0.33}$ & --- &67.68$^{+0.12}_{-0.12}$\\\noalign{\smallskip}
Position angle of the line of nodes, $\Omega$ [$^{\rm o}$] & 108.69$^{+0.22}_{-0.37}$ & --- &108.80$^{+0.14}_{-0.15}$\\\noalign{\smallskip}
Angular semimajor axis of the relative orbit, $a^{\prime\prime}$ [mas] & 24.71$^{+0.19}_{-0.16}$ & --- &25.15$^{+0.09}_{-0.08}$\\\noalign{\smallskip}
Systemic velocity, $\gamma$ [km\,s$^{-1}$] & --- & 9.61$^{+0.23}_{-0.21}$ & 9.59$^{+0.23}_{-0.21}$\\\noalign{\smallskip}
Half-range of radial velocity for primary, $K_1$ [km\,s$^{-1}$] & --- & 62.9$^{+0.6}_{-0.6}$ & 62.9$^{+0.6}_{-0.6}$\\\noalign{\smallskip}
Half-range of radial velocity for secondary, $K_2$ [km\,s$^{-1}$] & --- & 72.36$^{+0.31}_{-0.28}$ & 72.35$^{+0.30}_{-0.29}$\\\noalign{\smallskip}
\noalign{\smallskip}
\noalign{\smallskip}\hline 
\end{tabular}
\end{table*}

\begin{figure*}[!ht]
\resizebox{\hsize}{!}{\includegraphics{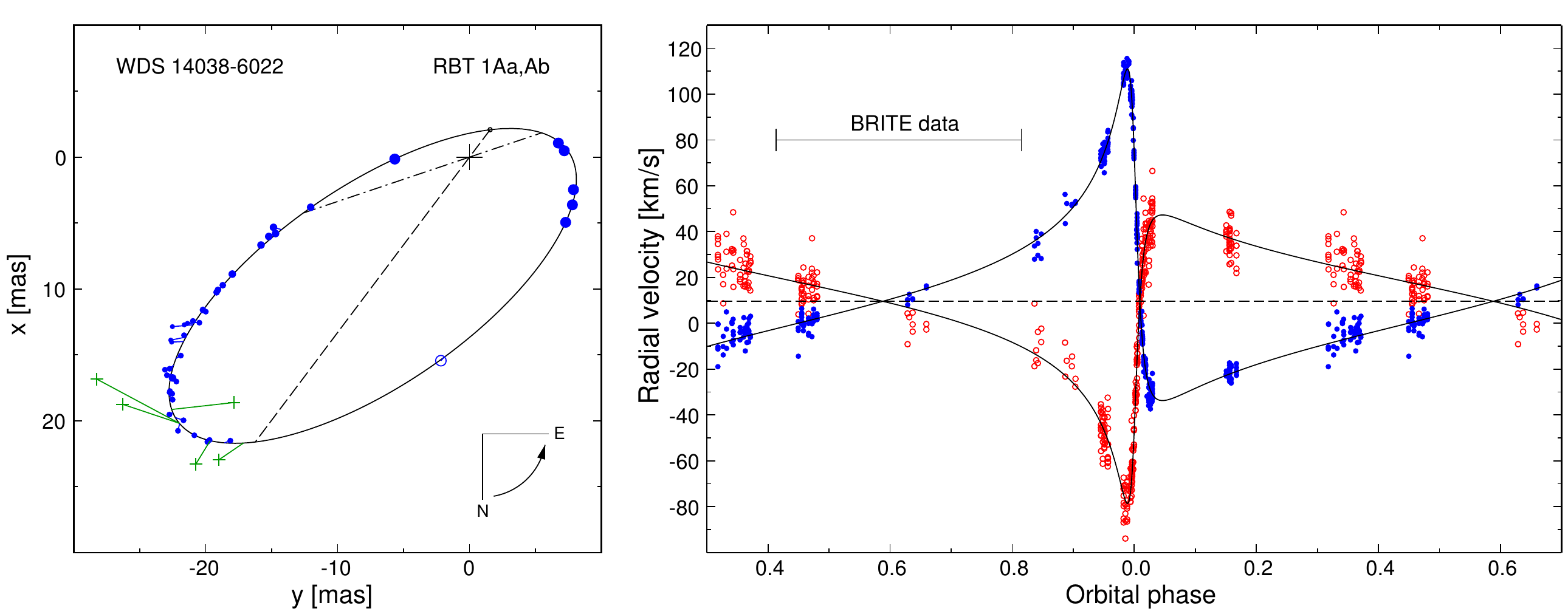}}
\caption{{\it Left}: Visual orbit of the Aa\,--\,Ab components of $\beta$~Cen. The MAPPIT observation is plotted as an open circle, speckle data with plus signs, SUSI observations with filled circles. For the last observations, the larger the symbol, the larger the weight that was adopted. The line of apsides is plotted with a dashed line, the line of nodes with a dash-dotted line. The designation of $x$ and $y$ follows the common double-star convention \citep{1978GAM....15.....H}. {\it Right}: Radial velocity curves for both components, filtered to remove any high-frequency contribution (red open circles for primary and blue filled circles for secondary), compared to the solution from combined visual and spectroscopic data (continuous lines). Phase 0.0 corresponds to $T_0$. The span of the B-C data is also shown.}
\label{AaAb-orbits}
\end{figure*}

\subsection{The orbit of Aa\,--\,Ab (Rbt\,1Aa,Ab)}\label{aaab}

The binarity of $\beta$~Cen\,A has been suspected owing to long-term variability of radial velocities \citep{1967MNRAS.136...51B} and intensity interferometer observations \citep{1974MNRAS.167..121H}. The orbital period of 352~d was first derived by \cite{1968PASAu...1...82S}. The Aa and Ab components were first resolved spatially with the MAPPIT interferometer \citep{1999MNRAS.302..245R}. Although \cite{1999MNRAS.302..245R} showed that the star is a double-lined spectroscopic binary, parameters of the spectroscopic orbit based on radial velocity curves of both components were first derived by \cite{2002A&A...384..209A}. They refined the orbital period to 357~d and found that the orbit is very eccentric ($e=$~0.81). It appeared that the components have very similar masses, but that one of them rotates significantly faster than the other. This observation was followed by several years of interferometric observations with the Sydney University Stellar Interferometer (SUSI) which enabled a derivation of all orbital parameters of the system, as well as the masses of the components \citep[both equal to 9.1 $\pm$ 0.3~$M_\odot$,][]{2005MNRAS.356.1362D}. A refined analysis of spectroscopic data which applied spectral disentangling \citep[hereafter Aus06]{2006A&A...455..259A} adjusted the mass ratio to about 0.88. It appeared that the component that rotates faster is more massive and consequently should be called the primary. This is opposite to the naming used previously. We therefore adopt the new convention proposed by Aus06 and will use it throughout this paper: the more massive and faster rotating component will be named the primary or Aa; its companion, secondary or Ab.

Since the study of Aus06, the Aa\,--\,Ab components were also resolved by means of speckle interferometry. Although the speckle observations \citep{2010AJ....139..743T,2014AJ....147..123T,2015AJ....150...50T,2012AJ....143...42H} are less accurate than those made with SUSI, the visual observations of the components now cover about 20 years and it is advisable to update the orbital parameters. The parameters of the visual orbit and their uncertainties were derived by means of a bootstrap method using a least-squares fit. With respect to the analysis of \cite{2005MNRAS.356.1362D}, we changed the weighting scheme for the SUSI observations: the weights were set equal to the baseline in meters as this better reflects the uncertainties of relative positions given by the authors. The MAPPIT and speckle observations were given weights equal to 1. The derived parameters of the visual orbit are given in the second column of Table \ref{aab-param}.\footnote{The epochs for the speckle data are Besselian \citep{1979A&A....73..282L}, so we translated all times for visual and radial velocity data to Besselian epochs. It appeared, however, that for the SUSI observations \citep{2005MNRAS.356.1362D} times translated from the original mean MJDs are late by about 3.5~d in the 4th Catalog of Interferometric Measurements. The same mistake is also reproduced in the WDS.}

The next step was to derive the parameters of the spectroscopic orbit. At the beginning, we obtained a preliminary spectroscopic solution using the 402 radial velocities published by Aus06 and available through the Centre de Donn\'ees Astronomiques de Strasbourg (CDS). The residuals from this fit were subjected to time-series analysis. The analysis revealed the presence of high frequencies that very likely come from pulsations, but also the presence of some instrumental/observational effects (e.g., frequency close to 1~d$^{-1}$). Unfortunately, strong aliasing prevents us from obtaining reliable frequencies for these data, despite knowing the frequencies from BRITE photometry (Sect.~\ref{br-phot}). The reason is that frequencies of the strongest peaks cannot be identified with aliases of frequencies that are derived from photometry. However, it is reasonable to remove them because this operation considerably reduces the scatter of the radial velocities in the orbital fit. On the other hand, the procedure does not affect the orbital radial velocity curve because the frequencies of the removed terms are much higher than the orbital frequency. In total, five periodic terms were derived and removed for the radial velocity variations of the primary and secondary. The radial velocities -- with the contributions of these high-frequency terms removed -- were subsequently used in the solution for the spectroscopic parameters. The parameters are given in the third column of Table \ref{aab-param}. The removal of high frequency terms lowered the residual standard deviation by 23 percent for radial velocities of the primary (to 7.1~km\,s$^{-1}$) and by 35 percent for the radial velocities of the secondary (to 3.6~km\,s$^{-1}$). As in the case of the visual orbit, the parameters and their uncertainties were derived by means of the bootstrap method. We note that the orbital period is much better constrained by the spectroscopic than by the visual data. This is a consequence of the radial velocity data covering two intervals in the vicinity of periastron passage, separated by two orbital periods, when the radial velocities change very rapidly (see Fig.~\ref{AaAb-orbits}). However, only one series of SUSI data was obtained that was close to periastron passage.

Since both types of orbit share four parameters, we finally combined the visual and spectroscopic data to obtain a common and final set of orbital parameters. The weighting scheme was the same as for the data analyzed separately. The final parameters of the combined visual and spectroscopic orbits are given in the fourth column of Table \ref{aab-param} and were derived in the same way as for the separate fits. The predicted orbital variations corresponding to the final parameters are compared with the visual and spectroscopic observations in Fig.~\ref{AaAb-orbits}.

The parameters allow masses and absolute magnitudes of the components to be determined as well as the size and distance of the system (Table \ref{massdist}). The masses are equal to $M_1=$ 12.02 $\pm$ 0.13~M$_\odot$ and $M_2=$ 10.58 $\pm$ 0.18~M$_\odot$ ($M_2/M_1=$~0.880 $\pm$ 0.008). This is at the upper limit of the uncertainties provided by Aus06. We also get slightly higher values of the size and distance of the system. The resulting parallax (9.04 $\pm$ 0.04 mas) is in fairly good agreement with the revised determination of the Hipparcos parallax of this star by \cite{2007A&A...474..653V}, 8.32 $\pm$ 0.50~mas.
\begin{table}[!ht]
\caption{Other parameters of $\beta$~Cen Aa\,--\,Ab system.}
\label{massdist}
\centering 
\begin{tabular}{lr@{ }l} 
\hline\hline\noalign{\smallskip}
Parameter & \multicolumn{2}{c}{Value} \\ 
\noalign{\smallskip}\hline\noalign{\smallskip}
Primary mass, $M_1$ & 12.02 &$\pm$ 0.13 M$_\odot$\\
Secondary mass, $M_2$ & 10.58 &$\pm$ 0.18 M$_\odot$\\
Semimajor axis, $a$ & 2.782 &$\pm$ 0.011 AU\\
Distance, $D$ & 110.6 &$\pm$ 0.5 pc\\
Parallax, $\pi$ & 9.04 &$\pm$ 0.04 mas\\
Primary absolute magnitude, $M_{V,1}$ & $-$4.03 &$\pm$ 0.10 mag\\
Secondary absolute magnitude, $M_{V,2}$ & $-$3.88 &$\pm$ 0.10 mag\\
\noalign{\smallskip}\hline 
\end{tabular}
\end{table}

The reddening of $\beta$~Cen is slight. In terms of the color excess, $E(B-V)$, it amounts to 0.028 $\pm$ 0.009~mag \citep{2005A&A...433..659N} as derived from IUE spectroscopy. Using Str\"omgren photometry \citep{1970AJ.....75..624C}, we get $E(b-y)=$ 0.026~mag which translates into $E(B-V)=$ 0.035~mag. Taking a simple mean of the two determinations of $E(B-V)$, we obtain a total absorption in the Johnson $V$ passband $A_V=$ 3.1$E(B-V)\approx$~0.10~mag. With $V=$~0.61~mag \citep{1966MNSSA..25...44C} and the distance from Table \ref{massdist} we get an absolute magnitude $M_V$ of the system equal to $-$4.71~mag. The magnitude difference between the components measured by \cite{2005MNRAS.356.1362D} in the blue (448~nm) passband amounts to 0.15~mag. A difference of 0.1\,--\,0.4~mag at slightly longer wavelengths was reported from speckle data. We note, however, that due to the $\pm$180$^{\rm o}$ ambiguity in the position angle in the interferometric data, it is not certain if the more massive component is the brighter one. Let us assume this to be the case. For two stars with masses equal to those given in Table \ref{massdist} put on the 10\,--\,15-Myr isochrone (cf.~Sect.~\ref{parmod}) we get a difference of about 0.5~mag between the components, using different evolutionary models. This is slightly larger than implied by the interferometric measurements. However, as shown, e.g., by \cite{2008ApJ...683..441G}, rapid rotation changes both the measured effective temperature and luminosity and is inclination-dependent. The primary component of $\beta$~Cen rotates with the projected rotational velocity of the order of 200\,--\,250 km\,s$^{-1}$ \citep{1999MNRAS.302..245R,2011A&A...536L...6A}. If its rotation axis were perpendicular to the orbital plane, this would lead to an apparent change of $\log(L/L_\odot)$ of about $-$0.14 (magnitude difference of 0.35~mag) with respect to a non-rotating star \citep{2008ApJ...683..441G}. Given the fact that the secondary rotates much more slowly, we conclude that there is no discrepancy between the observed magnitude difference of the components, their assumed coevality and the masses derived in Table \ref{massdist}.

Finally, we note that $\beta$~Cen Ab (the more slowly rotating secondary component of the SB2) was identified as a magnetic B star by \citet{2011A&A...536L...6A}. The strength and geometry of its magnetic field remain uncertain owing to the limited magnetic data (three observations) that are available. Nevertheless, it is reasonable to suspect that its magnetic nature has contributed to its slow (relative to the primary) rotation \citep[e.g.][]{2009MNRAS.392.1022U}.

\subsection{The orbit of A\,--\,B (Vou\,31AB)}
\begin{figure}[!ht]
\centering
\resizebox{0.8\hsize}{!}{\includegraphics{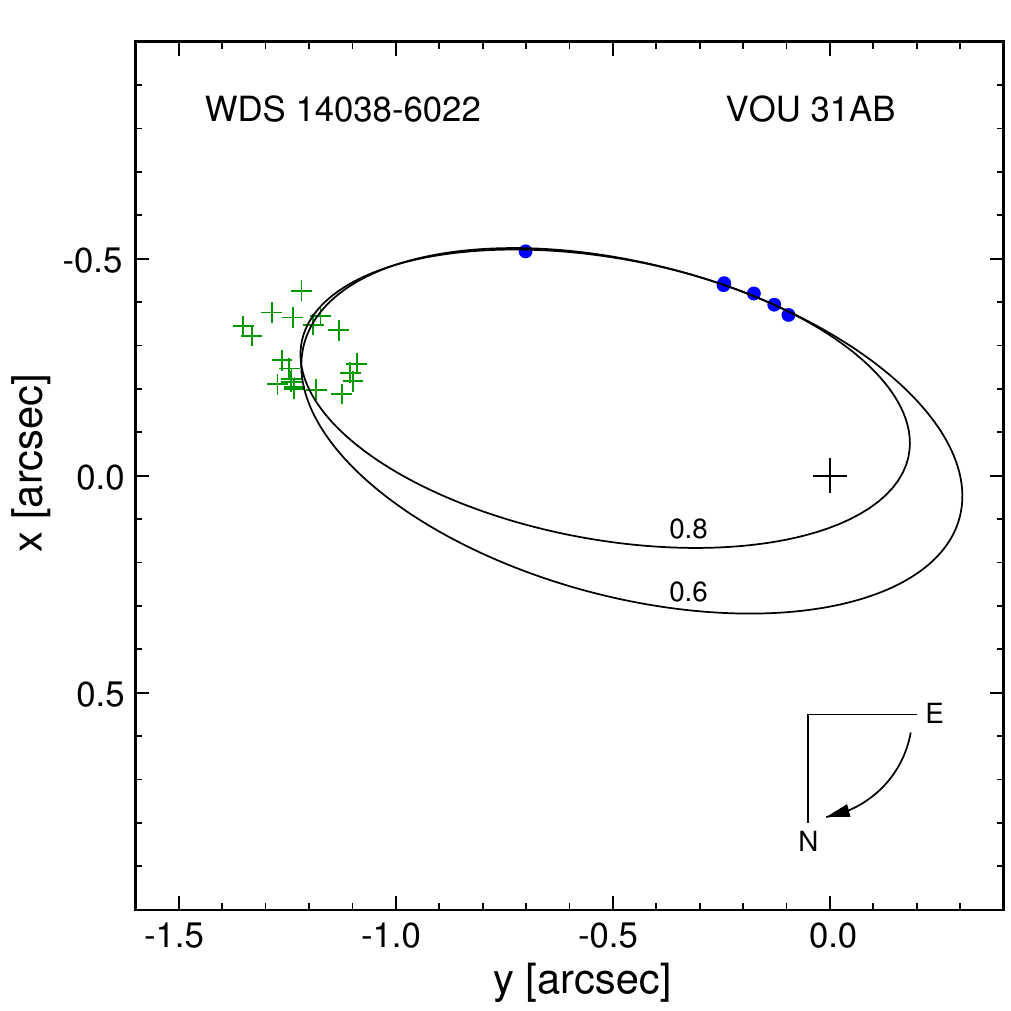}}
\caption{Visual orbit of the B component of $\beta$~Cen. Tycho and speckle observations are shown with filled circles, visual ones by plus signs. Two example orbits are labeled with their eccentricities.}
\label{AB-orbit}
\end{figure}

The wider A\,--\,B system is designated as Vou\,31AB after the discovery of the B component in 1935 by \cite{1947AnBos...6D...1V} at a separation of about 1.2~arcsec. The other visual measurements of the A\,--\,B pair were obtained by \cite{1936CiUO...96..263F,1941CiUO..104...74F}, \cite{1937CiUO...98..362V,1947CiUO..105..134V,1956CiUO..115..266V,1957CiUO..116R.291V,1961CiUO..120..353V}, \cite{1948MmMtS...9....1W}, \cite{1951AnBos...9.....S}, \cite{1955POMic..11....1R}, \cite{1955JO.....38..109V}, and \cite{1956MNRAS.116..248H}. The components were also resolved by Tycho \citep{2002A&A...384..180F}. New observations made by means of speckle interferometry \citep{2010AJ....139..743T,2014AJ....147..123T,2015AJ....150...50T,2012AJ....143...42H} indicate much smaller and rapidly decreasing separation between the components (Fig.~\ref{AB-orbit}). The arc is too short to allow the orbital parameters to be determined, but they can be constrained. We made many fits assuming different values for the eccentricity. Prior to this, the positions of the B component from speckle data (in which Aa and Ab are resolved) were corrected for the photocenter of the A component. This was done under the assumption that the flux ratio is equal to 0.868 \citep{2005MNRAS.356.1362D} and the brighter component is Aa. The constraints on the parameters are the following: $e>$~0.5 and formally can be as large as 0.99. However, it is rather unlikely that $e$ is so high because it would lead to strong dynamical interactions with the inner pair. We therefore speculate that $e$ is probably not higher than 0.9. On the other hand, $e$ is likely not smaller than 0.5 because orbits with $e<$~0.5 do not fit well the recent, most accurate speckle data. Given the deduced range of eccentricities, 0.5\,--\,0.9, the following constraints on the remaining parameters can be inferred: $P_{\rm orb}=$~125\,--\,220 yr, $\omega=$~150$^{\rm o}$\,--\,240$^{\rm o}$, $T_0=$~2024\,--\,2032, $\Omega=$~67$^{\rm o}$\,--\,110$^{\rm o}$, $a^{\prime\prime} =$~0.75\,--\,1.0~arcsec, $i=$~118$^{\rm o}$\,--\,130$^{\rm o}$. The orbits for two values of eccentricity (0.6 and 0.8) are shown in Fig.~\ref{AB-orbit}. It is clear that observations obtained over the next decade should provide the full solution of this orbit.

\section{Photometry from BRITEs}\label{br-phot}
\begin{figure}[!ht]
\resizebox{\hsize}{!}{\includegraphics{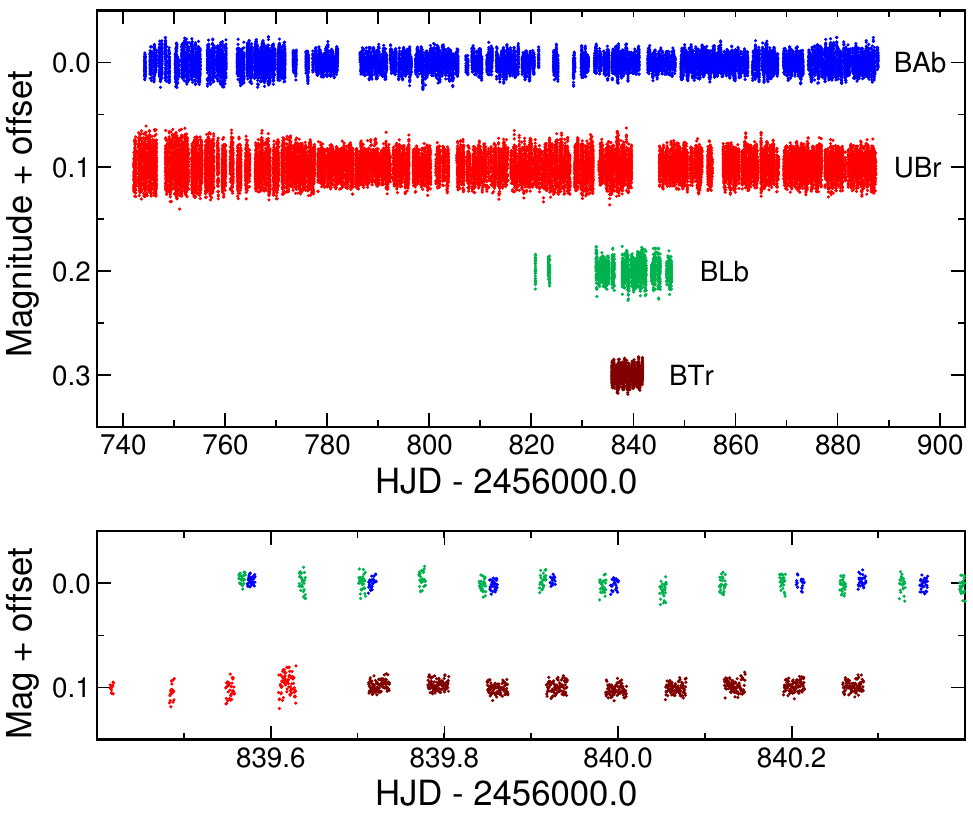}}
\caption{{\it Top:} Distribution of BRITE data of $\beta$~Cen from four BRITE satellites. Offsets have been applied to separate the light curves. {\it Bottom:} A 1-day series of BRITE observations of $\beta$~Cen. Note groups of data points separated by the satellites' orbital period of about 100~minutes (0.07~d). The same offsets have been applied to the blue-filter (BAb + BLb, upper sequence of points) and red-filter (UBr + BTr, lower sequence of points) data.}
\label{B-data}
\end{figure}

The star $\beta$~Cen was one of 32 observed by B-C in the Centaurus field, the second target field of B-C following Orion (the commissioning field for the two Austrian BRITEs). The observations spanned nearly 146 days between March 25 and August 18 2014 and come from four BRITEs: blue-filter BRITE-Austria (hereafter BAb) and Lem (BLb) and red-filter Uni-BRITE (UBr) and BRITE-Toronto (BTr). Most observations were obtained with BAb and UBr. They were supplemented by a 27-day run of BLb and a 6-day run of BTr; see Fig.~\ref{B-data}. Details of the instrumentation and observing procedure are given by \cite{2014PASP..126..573W}.

Due to the limitations of telemetry, no full images are downloaded from BRITEs. Instead, rasters around selected bright stars are stored in the onboard memory and downloaded to ground stations during nearby passes. The images are intentionally defocused to avoid saturation and to decrease the impact of pixel-to-pixel sensitivity variations on the final photometry. The current reduction pipeline, developed by Adam Popowicz, is based on aperture photometry with constant, circular apertures that are optimised for each star, satellite, and observational setup. The only correction to the measured fluxes at this step is for intra-pixel variability, which is approximated by a fourth-order polynomial. Details of the reduction procedure will be published by Popowicz et al.~(in preparation).

The raw B-C photometry which comes from the reduction pipeline is not suitable for frequency analysis. It requires a series of preparatory steps which are explained in detail in Appendix \ref{prep}. This procedure ultimately yields two light curves, one for the blue BRITE passband, and the other for the red passband.

\subsection{Time-series analysis of the B-C data}\label{tser}
The blue- and red-filter B-C light curves consist of 40\,293 and 63\,214 data points, respectively. Each data point is the result of a 1-second exposure. For many reasons the data are of different quality. The median standard deviations, estimated using orbit-averaged samples, equal 5.0~mmag for BAb, 6.2~mmag for BLb, 8.5~mmag for UBr, and 4.7~mmag for BTr. To search for variability of $\beta$~Cen in B-C photometry, the blue- and red-filter data were combined together. Since amplitudes in the red filter are smaller (at least part of this effect can be attributed to instrumental effects, see Appendix A), the red-filter light curve was multiplied by a factor of 1.5 prior to combining. This factor was derived as a mean of amplitude ratios for the five strongest terms fitted separately to the blue- and red-filter light curves. There are two main advantages to combining the data. First, this lowers the detection threshold, which may allow low-amplitude modes, which are not seen in single-filter data, to be detected. Next, since the orbital periods of the BRITE satellites are slightly different, the data from one satellite can fill gaps in the observations obtained by the others; see Fig.~\ref{B-data}, bottom panel. As a consequence, combining data considerably reduces aliases and helps to identify the true frequencies unambiguously.

\begin{figure*}
\sidecaption
\includegraphics[width=12cm]{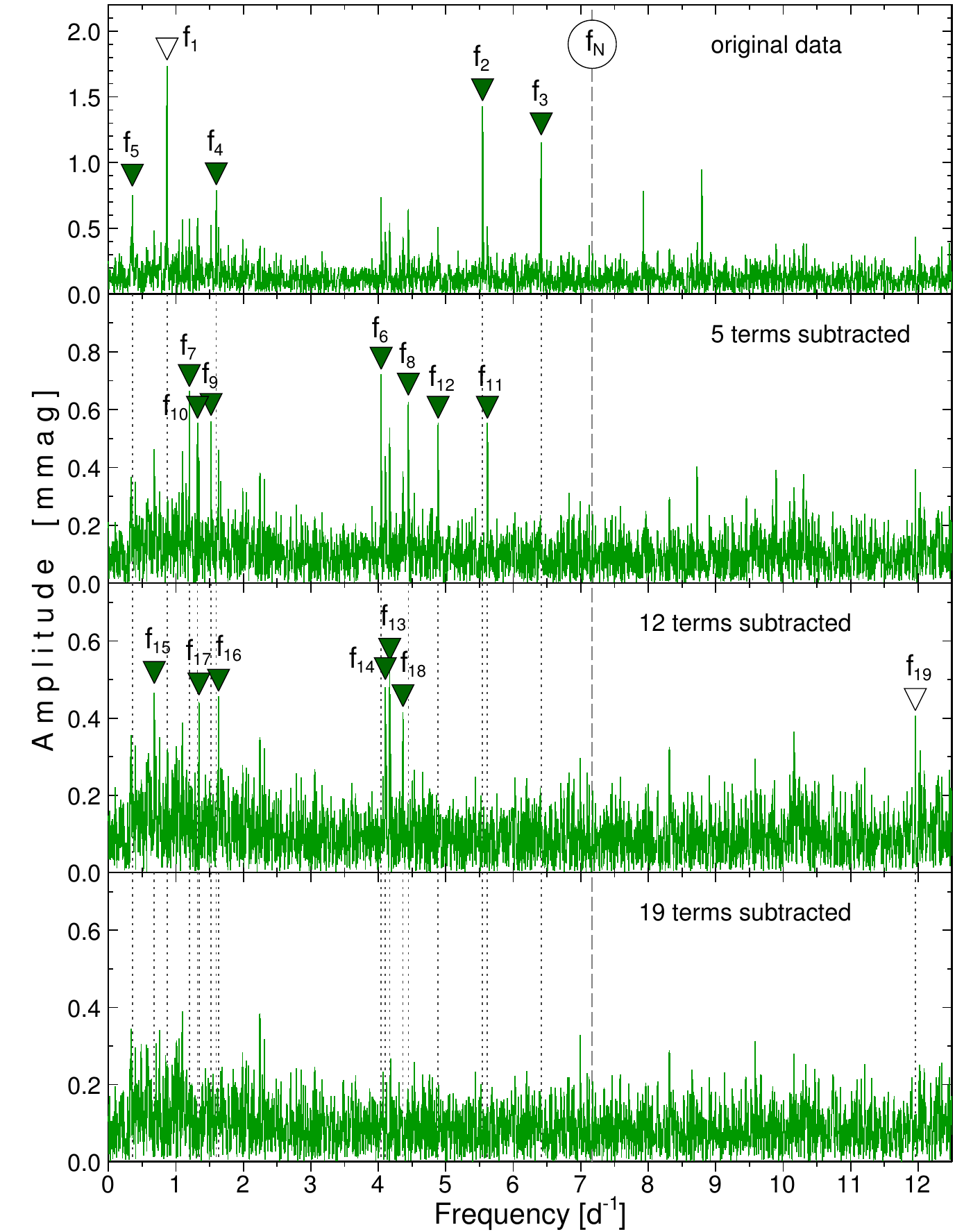}
\caption{Fourier frequency spectra for the combined blue- and red-filter B-C data of $\beta$~Cen at four steps of the pre-whitening procedure. The detected terms are marked with inverted triangles and labeled from $f_1$ to $f_{19}$. The open triangles for $f_1$ and $f_{19}$ indicate combination frequencies. $f_{\rm N}\approx$ 7.17~d$^{-1}$ stands for the Nyquist frequency related to the orbital sampling and is marked with a dashed line. The dotted vertical lines indicate frequencies of terms that were subtracted. Note the different ordinate scales in different panels.}
\label{fs-cmb}
\end{figure*}
\begin{figure*}
\sidecaption
\includegraphics[width=12cm]{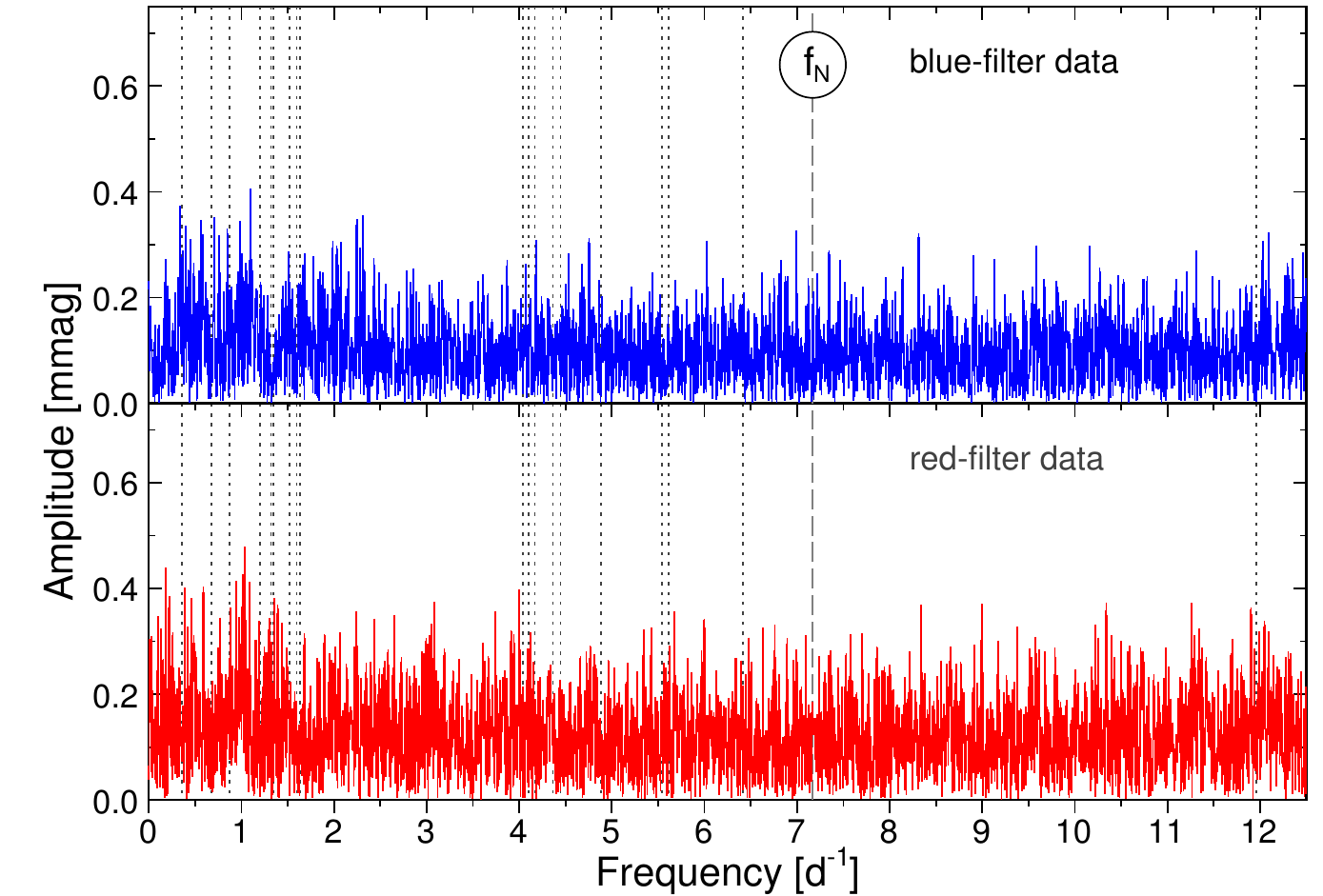}
\caption{Fourier frequency spectra for residuals from the 19-term fit to the blue-filter (top) and red-filter (bottom) B-C data of $\beta$~Cen. The dotted and dashed lines have the same meaning as in Fig.~\ref{fs-cmb}.}
\label{fs-resid}
\end{figure*}

\begin{table*}[!ht]
\caption{Parameters of the least-squares fit to the B-C blue-filter (index $B$) and red-filter (index $R$) data. The phases are given at epoch HJD\,2456840.0. The last column includes linear rates of frequency changes, $\beta$; see Sect.~\ref{aaorab} for explanation.}
\label{freqsol}
\centering\small
\begin{tabular}{cr@{.}lcccr@{.}lr@{.}lr@{.}l} 
\hline\hline\noalign{\smallskip}
& \multicolumn{2}{c}{Frequency} & \multicolumn{2}{c}{Semi-amplitudes} & \multicolumn{5}{c}{Phases and phase difference} & \multicolumn{2}{c}{$\beta_i$}\\ 
$f_i$ & \multicolumn{2}{c}{[d$^{-1}$]}& $A_B$ [mmag] & $A_R$ [mmag] & $\phi_B$ [rad] & \multicolumn{2}{c}{$\phi_R$ [rad]} & \multicolumn{2}{c}{$\phi_B-\phi_R$ [rad]}&\multicolumn{2}{c}{[10$^{-6}$\,d$^{-2}$]}\\ 
\noalign{\smallskip}\hline\noalign{\smallskip}
$f_1= f_3-f_2$ & 0&86833 & 1.811 $\pm$ .036 & 1.220 $\pm$ .045 & 0.267 $\pm$ .023 & 0&146 $\pm$ .038 & $+$0&12 $\pm$ .05&\multicolumn{2}{c}{---}\\
$f_2$ & 5&54517 $\pm$ .00009 & 1.540 $\pm$ .036 & 0.786 $\pm$ .046 & 0.125 $\pm$ .025 & $-$0&018 $\pm$ .066 & $+$0&14 $\pm$ .07 &$-$1&8$^{+2.2}_{-2.3}$\\
$f_3$ & 6&41350 $\pm$ .00010 & 1.069 $\pm$ .038 & 0.554 $\pm$ .048 & 1.777 $\pm$. 035 & 1&673 $\pm$ .086 & $+$0&10 $\pm$ .09 &+6&7$^{+4.2}_{-4.1}$\\
$f_4$ & 1&59815 $\pm$ .00023 & 0.680 $\pm$ .039 & 0.405 $\pm$ .047 & 0.776 $\pm$ .058 & 0&465 $\pm$ .126 & $+$0&31 $\pm$ .14 &+1&9$^{+5.8}_{-5.0}$\\
$f_5$ & 0&35367 $\pm$ .00019 & 0.609 $\pm$ .036 & 0.547 $\pm$ .047 & 4.274 $\pm$ .061 & 4&092 $\pm$ .087 & $+$0&18 $\pm$ .11&$-$1&8$^{+3.7}_{-4.8}$\\
$f_6$ & 4&04180 $\pm$ .00019 & 0.771 $\pm$ .040 & 0.694 $\pm$ .047 & 4.757 $\pm$ .051 & 4&350 $\pm$ .069 & $+$0&41 $\pm$ .09 &$-$8&5$^{+3.0}_{-3.5}$\\
$f_7$ & 1&19867 $\pm$ .00018 & 0.717 $\pm$ .039 & 0.448 $\pm$ .047 & 3.462 $\pm$ .052 & 3&563 $\pm$ .108 & $-$0&10 $\pm$ .12 &$-$8&7$^{+4.2}_{-3.9}$\\
$f_8$ & 4&44439 $\pm$ .00029 & 0.507 $\pm$ .038 & 0.331 $\pm$ .045 & 2.937 $\pm$ .080 & 3&179 $\pm$ .141 & $-$0&24 $\pm$ .16 &$-$9&3$^{+4.4}_{-3.9}$\\
$f_9$ & 1&52100 $\pm$ .00028 & 0.679 $\pm$ .039 & 0.291 $\pm$ .048 & 1.032 $\pm$ .058 & 0&685 $\pm$ .164 & $+$0&35 $\pm$ .18 &+3&1$^{+4.2}_{-4.2}$\\
$f_{10}$ & 1&32022 $\pm$ .00030 & 0.553 $\pm$ .037 & 0.501 $\pm$ .046 & 2.595 $\pm$ .072 & 2&330 $\pm$ .097 & $+$0&26 $\pm$ .12 &+15&4$^{+3.7}_{-4.9}$\\
$f_{11}$ & 5&61725 $\pm$ .00040 & 0.492 $\pm$ .037 & 0.318 $\pm$ .048 & 1.288 $\pm$ .077 & 0&895 $\pm$ .151 & $+$0&39 $\pm$ .17 &+28&8$^{+7.3}_{-9.9}$\\
$f_{12}$ & 4&88617 $\pm$ .00023 & 0.563 $\pm$ .039 & 0.419 $\pm$ .047 & 0.122 $\pm$ .067 & $-$0&090 $\pm$ .111 & $+$0&21 $\pm$ .13 &$-$1&8$^{+6.5}_{-7.7}$\\
$f_{13}$ & 4&16793 $\pm$ .00021 & 0.572 $\pm$ .038 & 0.312 $\pm$ .050 & 5.698 $\pm$ .068 & 5&471 $\pm$ .169 & $+$0&23 $\pm$ .18 &$-$14&9$^{+6.7}_{-7.0}$\\
$f_{14}$ & 4&10357 $\pm$ .00045 & 0.424 $\pm$ .038 & 0.356 $\pm$ .049 & 3.218 $\pm$ .092 & 2&875 $\pm$ .134 & $+$0&34 $\pm$ .16 &$-$2&2$^{+6.1}_{-6.2}$\\
$f_{15}$ & 0&67907 $\pm$ .00032 & 0.371 $\pm$ .037 & 0.250 $\pm$ .048 & 0.247 $\pm$ .101 & $-$0&108 $\pm$ .196 & $+$0&36 $\pm$ .22 &$-$11&6$^{+11.5}_{-8.3}$\\
$f_{16}$ & 1&63040 $\pm$ .00049 & 0.381 $\pm$ .036 & 0.203 $\pm$ .047 & 0.607 $\pm$ .098 & 0&291 $\pm$ .243 & $+$0&32 $\pm$ .26 &+16&9$^{+42.0}_{-38.7}$\\
$f_{17}$ & 1&34157 $\pm$ .00027 & 0.425 $\pm$ .036 & 0.308 $\pm$ .048 & 4.913 $\pm$ .090 & 5&404 $\pm$ .168 & $-$0&49 $\pm$ .19 &$-$2&4$^{+8.9}_{-11.6}$\\
$f_{18}$ & 4&36444 $\pm$ .00035 & 0.262 $\pm$ .036 & 0.391 $\pm$ .049 & 1.131 $\pm$ .148 & 1&262 $\pm$ .123 & $-$0&13 $\pm$ .19 &$-$13&2$^{+38.6}_{-20.6}$\\
$f_{19} = f_2+f_3$ & 11&95867 & 0.364 $\pm$ .036 & 0.354 $\pm$ .046 & 3.549 $\pm$ .106 & 3&461 $\pm$ .137 & $+$0&09 $\pm$ .18 &\multicolumn{2}{c}{---}\\
\noalign{\smallskip}\hline 
\end{tabular}
\end{table*}

The resulting combined light curve was then used for time-series analysis, which consisted of the calculation of the Fourier frequency spectrum, identification of the highest maximum in the spectrum, and pre-whitening the original light curve with all previously detected frequencies. To account for variable quality of the data, summing terms in the frequency spectrum formula were weighted with the reciprocal of the squared standard deviation. The latter was estimated from the scatter in orbit samples in the preliminary solution separately for each set of satellite data. The Fourier frequency spectra at several stages of the pre-whitening procedure are shown in Fig.~\ref{fs-cmb}. In total, 18 significant terms were identified with signal-to-noise ratio (S/N) exceeding 4, the limiting value for our pre-whitening procedure. The terms are numbered in order of detection. We decided to add a 19th term which occurred with S/N $=$ 3.86 only because it is a combination frequency, so there is little doubt that it is real. The improved frequencies and their uncertainties were subsequently derived using the combined data by means of the bootstrap method. These are given in Table \ref{freqsol} and can be identified in Fig.~\ref{fs-cmb}.

Of the 19 detected periodic terms, 17 can be identified as pulsation modes, nine in the low-frequency regime, where gravity ($g$) modes occur and eight in the high-frequency region where pressure ($p$) modes are observed. It is worth noting, however, that the possibility that one or two low-frequency terms correspond to the rotational frequency of the primary or secondary, cannot be excluded. This is more likely for the secondary, in which a magnetic field has been discovered by \cite{2011A&A...536L...6A}. The observed frequency spectrum makes $\beta$~Cen a hybrid $\beta$~Cep/SPB star and one of the few stars of this type with a rich frequency spectrum.

The two remaining terms, $f_1$ and $f_{19}$, are combination terms. In pulsating stars combination terms usually occur with amplitudes much smaller than the parent modes. Here, the situation is different because $f_3$ = $f_1 + f_2 =$ 6.41350~d$^{-1}$ has an amplitude comparable to that of $f_1$ and $f_2$. Therefore, a possibility occurs that of the three terms, $f_1$, $f_2$, and $f_3$, any two could be intrinsic and the remaining one, a combination. The other combination term we detected is $f_{19} = f_1 + \mbox{2}f_2$. The occurrence of such a combination term is rather surprising because usually the lowest order combination terms occur first. Assuming that the intrinsic modes are $f_2$ and $f_3$, both $f_1 = f_3 - f_2$ and $f_{19} = f_2 + f_3$ can be interpreted as the lowest-order combinations. Our assumption also leads to the conclusion that the combination frequency ($f_1$) has the largest amplitude of all detected terms. This is rather unusual, but the possibility exists and can be even explained theoretically \citep{2015MNRAS.450.3015K}. With this assumption, our solution consists of 8 $g$ and 9 $p$ modes. Consequently, the solution in Table \ref{freqsol} is presented under this assumption. We note that previous determinations of high-frequency variations in spectroscopic data revealed only $f_2$ and/or $f_3$ (or their aliases); see Sect.~\ref{SMEI}. 

Having derived frequencies from the combined data set, we fixed them and fitted the 19-term model to the blue and red B-C data separately. The resulting semi-amplitudes and phases are listed in Table \ref{freqsol}, while the Fourier spectra of the residuals from the fit in each filter are shown in Fig.~\ref{fs-resid}.

\subsection{Comparison with previous results}\label{SMEI}
The short-period radial velocity variations found in $\beta$~Cen by \cite{1967MNRAS.136...51B} and confirmed in other studies \citep{1968PASAu...1...82S,1975MNRAS.172..639L,1982AcA....32..371K,1999MNRAS.302..245R,2002A&A...384..209A,2006A&A...455..259A} were attributed to $\beta$~Cep-type pulsations of one or both bright components. Unfortunately, the presence of two relatively rapidly rotating components have historically complicated the extraction of the correct frequencies. A summary of the frequencies attributed to pulsations that were reported in the literature is given in the top part of Table \ref{frid}. It can be seen that they correspond to one of the two strongest $p$ modes ($f_2$ or $f_3$) or their daily (and monthly) aliases.

\begin{table}[!ht]
\caption{Identification of the short-period variability of $\beta$~Cen found in the spectroscopy (top part) and photometry (bottom part).}
\label{frid}
\centering
\begin{tabular}{llll} 
\hline\hline\noalign{\smallskip}
 & Freq. & \\ 
Source & [d$^{-1}$] & Identification \\ 
\noalign{\smallskip}\hline\noalign{\smallskip}
\cite{1967MNRAS.136...51B} & 7.59 & $f_2+\mbox{2~d}^{-1}$\\
\cite{1968PASAu...1...82S} & 7.418 & $f_3+\mbox{1~d}^{-1}$\\
\cite{1975MNRAS.172..639L} & 6.37 & $f_3$\\
\cite{1999MNRAS.302..245R} & 5.59 & $f_2$\\
& 6.28 & possibly $f_3$\\
Ausseloos et al.~(2002) & 6.51481 & $f_2+\mbox{1~d}^{-1}$\\
&& \quad$-\mbox{1~mo}^{-1}$\\
& 6.41356 & $f_3$ \\
& 6.49521 & --- \\
Ausseloos et al.~(2006) & 7.415 & $f_3+\mbox{1~d}^{-1}$\\
& 4.542 & $f_2-\mbox{1~d}^{-1}$\\
\noalign{\smallskip}\hline\noalign{\smallskip}
\cite{1977MmRAS..84..101B} & 3.33 & ---\\
\cite{2012PhDT..Goss} & 0.86796 & $f_1$\\
& 5.54508 & $f_2$\\
& 6.41313 & $f_3$\\
& 4.88538 & $f_{12}$\\
& 1.19941 & $f_7$\\
& 4.44438 & $f_8$\\
& 1.28907 & ---\\
& 2.55313 & ---\\
& 1.3203 & $f_{10}$\\
& 2.51893 & ---\\ 
\noalign{\smallskip}\hline 
\end{tabular}
\end{table}

Ground-based time-series photometry of $\beta$~Cen is very sparse. \cite{1973MNRAS.162...25S} did not find any variations, \cite{1977MmRAS..84..101B} found the star to be variable on one night and constant on the other. \cite{1982AcA....32..371K} detected only a small level of variability in the Str\"omgren $u$ band. Quite recently, photometry of $\beta$~Cen was obtained by the SMEI experiment on-board the Coriolis satellite \citep{2004SoPh..225..177J} and analysed by \cite{2012PhDT..Goss}. She finds ten significant terms, of which we recover seven; see the bottom part of Table \ref{frid}.

\subsection{Aa or Ab: that is the question}\label{aaorab}
Since both the Aa and Ab components are early B-type stars, the pulsations we detected can be excited in either of them. In principle, the $g$ modes can be excited even in the B component which, given the 3- to 4-mag difference with respect to the A component, should be a mid B-type star. Analysing the variability of the profile of the Si\,{\sc iii} 4553~{\AA} line, \cite{1999MNRAS.302..245R} attributed pulsations to the Ab component. However, they found the same two frequencies in the variability of the primary's wider lines which, in view of strong blending, rather leads to the conclusion that the pulsations have their origin in the Aa component. \cite{1999MNRAS.302..245R} also indicated that line profiles of both components vary, a result later that is confirmed by \cite{2002A&A...384..209A}. A thorough analysis performed by Aus06 shows that both the pulsation modes that they found (aliases of $f_2$ and $f_3$; see Table \ref{frid}) should, in fact, be attributed to the Aa component.
\begin{figure}
\centering
\includegraphics[width=0.45\textwidth]{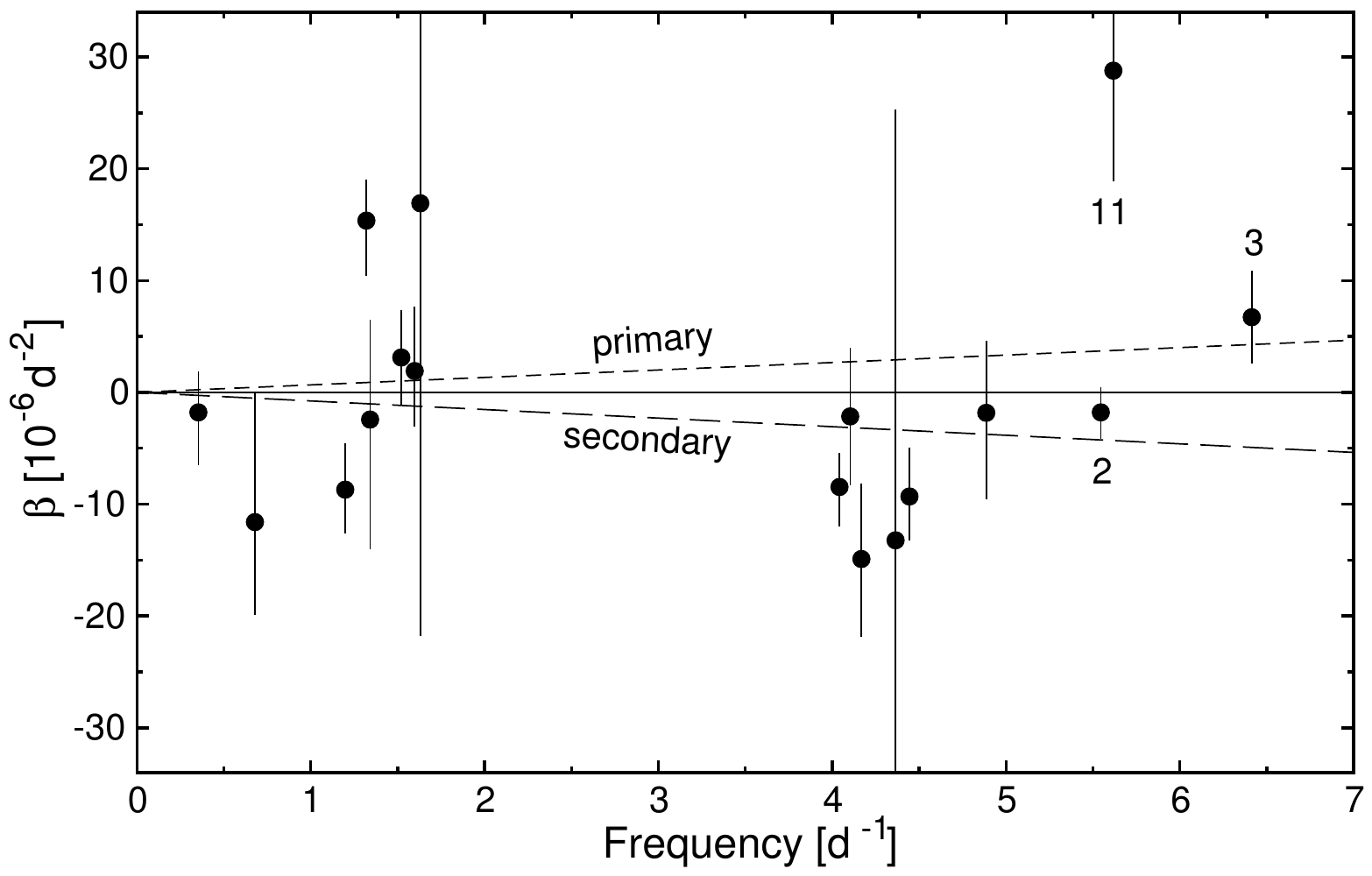}
\caption{Derived linear rates of frequency change, $\beta$, for 17 intrinsic modes found in the B-C data of $\beta$~Cen plotted as a function of their frequencies. For comparison, the expected rates for the primary (short-dashed line) and secondary (long-dashed line) are shown. Modes discussed in the text are labeled with their numbers from Table \ref{freqsol}.}
\label{rates}
\end{figure}

There is, however, a possibility to verifying this finding and perhaps attribute other modes to the components using the B-C photometry. As can be seen from Fig.~\ref{AaAb-orbits}, the B-C data cover orbital phases between 0.413 and 0.815. In this phase interval, the radial velocity of the primary (Aa) changes almost linearly at a rate of about $\alpha_{\rm Aa}=-$0.20\,km\,s$^{-1}$\,d$^{-1}$ whereas that of the secondary (Ab) changes at a rate of $\alpha_{\rm Ab}=+$0.23\,km\,s$^{-1}$\,d$^{-1}$. The Doppler effect should therefore lead to the increase of frequencies of all modes that were excited in the primary at a rate of $\beta_{\rm Aa}\approx-\alpha_{\rm Aa}f_{\rm intr}/c$, where $f_{\rm intr}$ is the intrinsic frequency of a mode and $c$ is the speed of light. Similarly, the modes excited in the secondary should decrease their frequencies at a rate of $\beta_{\rm Ab}\approx-\alpha_{\rm Ab}f_{\rm intr}/c$. Using the combined blue and red B-C data set, we allowed for linear frequency changes and derived the rates $\beta$ for all intrinsic modes. The rates are reported in Table \ref{freqsol} and shown in Fig.~\ref{rates}. 

It can be seen from Fig.~\ref{rates} that the uncertainties of $\beta$, derived by means of a bootstrap method, are large in comparison with the expected effect. Therefore, no reliable conclusion as to the component(s) in which $g$ modes are excited can be drawn. This is understandable because the B-C data span too short a time interval to get $\beta$ with a sufficient precision for $g$ modes. The situation is slightly better for $p$ modes which have higher frequencies and, consequently, larger |$\beta$| values are expected for them (since $|\beta|$ is proportional to frequency). The value of $\beta$ for the $f_3$ mode indicates the primary component as the origin of this mode, in agreement with Aus06. The same should apply to $f_2$. Surprisingly, $\beta_2=-$1.8$^{+2.2}_{-2.3}$\,$\times$\,10$^{-6}$\,d$^{-2}$ is negative and fits better the expected rate for the secondary (within 1.1$\sigma$) than the primary (within 2.5$\sigma$). On the other hand, we know that $f_2$ and $f_3$ have to be excited in the same component, since they form combination frequencies. This leads us to the conclusion that it is probably the primary in which $f_2$ and $f_3$ are excited. The slightly too low value of $\beta_2$ that we get can probably be explained by the presence of a nearby mode, $f_{11}$.

As far as the origin of the other $p$ modes is concerned, their $\beta$ values are negative, indicating the secondary as the origin. Given the uncertainties of the $\beta$s, however, the possibility that some of them are excited in the primary cannot be excluded. In summary, the idea of using rates of frequency change to separate modes in both components in $\beta$~Cen was of limited use. However, there is the possibility that the Centaurus field will be re-observed by B-C in the future. With another season, the rates $\beta$ could be derived more precisely which, at least for the $p$ modes, may provide a conclusive result as to their origin.

\section{Asteroseismic aspects}\label{smodel}
The most important application of the measured oscillation spectrum is deriving constraints on a stellar structure by the construction of seismic models whose oscillation frequencies best reproduce those observed. The task is complicated by the requirement of mode identification (hereafter MID) in terms of three quantum numbers: angular degree, $l$, angular order, $m$, and radial order, $n$, of a spherical harmonic, which defines a given mode of pulsation. MID is typically achieved with the use of photometric and radial velocity amplitudes and phases as well as variability of line profiles \citep[see][for a recent review of MID methods]{2014IAUS..301...93B,2014IAUS..301..101U}. Unfortunately, the instrumental effects (explained in Appendix A) in $\beta$~Cen BRITE data very likely affect the amplitudes in both passbands. This means that it is better not to use the $A_B/A_R$ amplitude ratios for MID in this star; see also Sect.~\ref{discon}.

With 17 observed intrinsic modes of oscillation occuring in both $p$- and $g$-mode frequency domains, $\beta$~Cen seems to be particularly suitable for asteroseismic modeling. The most severe complication in seismic modeling of this star is the rapid rotation of the components and the inability to assign the observed modes to particular components of the binary unambiguously. Therefore, no final seismic model of the star is given here. However, to illustrate the most important effects that need to be taken into account in modeling this star, we provide a discussion of stellar parameters of the components followed by a review of the effects of rotation and rotational coupling. This consideration may provide a good starting point for building seismic models of the components once observed modes have been assigned to the components and identified.

\subsection{Stellar parameters and models}\label{parmod}
We adopted the derived frequencies (Table \ref{freqsol}) and stellar masses from Table \ref{massdist} as the primary input parameters for the model calculations. The bright components show markedly different rotational velocities; the more massive primary component rotates faster than the secondary. In some previous determinations of the projected rotational velocities the components were not resolved, so that the values of $V_{\rm rot}\sin i$ derived by \cite{1964IzKry..31...44B}, \cite{1970PASP...82..741L}, \cite{1972ApJS...24..167W}, and \cite{1975ApJS...29..137S}: 128, 175, 105 and 70~km\,s$^{-1}$, respectively, diverge because they probably correspond either to some kind of a mean or the value for the secondary. \cite{1999MNRAS.302..245R} provided upper limits of 
265 and 120~km\,s$^{-1}$ for $V_{\rm rot}\sin i$ of the primary and secondary, while \cite{2011A&A...536L...6A} report 190 $\pm$ 20 and 75 $\pm$ 15~km\,s$^{-1}$. Assuming that the rotation axes are perpendicular to the orbital plane, we find that the primary's $V_{\rm rot}$ is of the order of 200\,--\,250~km\,s$^{-1}$, while that of the secondary amounts to about half of that, 70\,--\,120~km\,s$^{-1}$.

\begin{figure*}
\centering
\includegraphics[angle=270,width=0.6\textwidth]{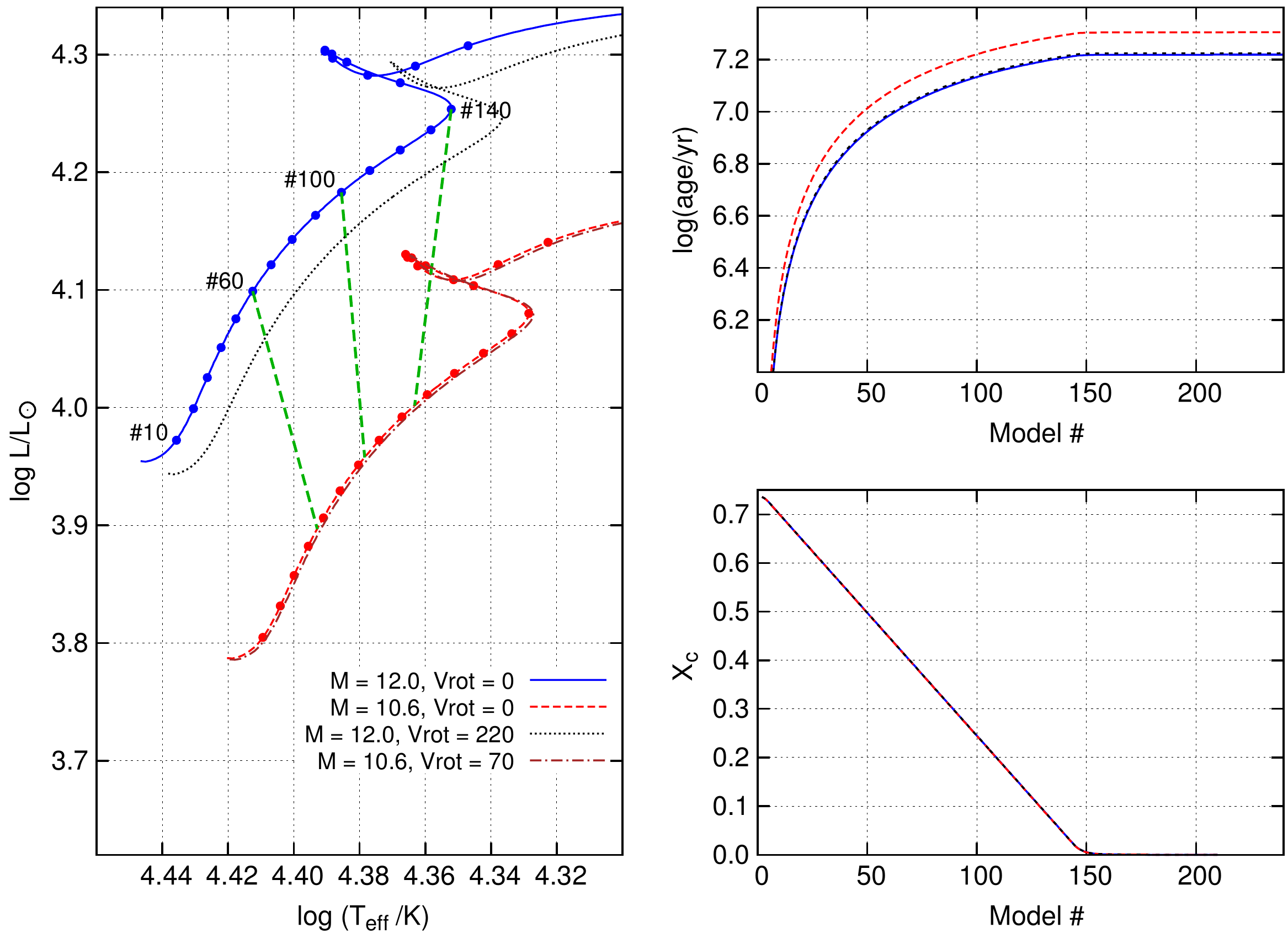}
\caption{\textit{Left}: Evolutionary tracks for stellar models with masses equal to 12.0 and 10.6 M$_\odot$ and different values of $V_{\rm rot}$ (see the labels). The OP opacities and initial composition of $X =$ 0.7374 and $Z =$ 0.0134 were adopted. Every tenth model for the non-rotating case is indicated with a dot, starting from model \#10 (labeled). Three coeval pairs of models are connected with dashed lines and labeled with model number in the 12.0-M$_\odot$ evolutionary track; see text for explanation. \textit{Right}: Age (top) and central hydrogen mass abundance, $X_{\rm c}$, (bottom) plotted as a function of model number. The relation for 12.0\,M$_\odot$ is plotted with a continuous line, and for 10.6\,M$_\odot$, with a dashed line.}
\label{BCen20a}
\end{figure*}

As far as the effective temperatures, $T_{\rm eff}$, are concerned, it is usually noted that the spectra of both bright components of $\beta$~Cen are similar and a single value of $T_{\rm eff}$ is derived. This is not a surprise, given the fact that, if the stars are coeval, they should have very similar effective temperatures (see Fig.~\ref{BCen20a}, left panel). From the calibration of Str\"omgren photometry, \cite{1993SSRv...62...95S} derived $\log (T_{\rm eff}/\mbox{K}) =$ 4.374 ($T_{\rm eff}=$ 23\,700~K). From the fit to the IUE spectra, \cite{2005A&A...433..659N} get a slightly higher value, $\log (T_{\rm eff}/\mbox{K}) =$ 4.408\,$\pm$\,0.020 ($T_{\rm eff}=$ 25\,600\,$\pm$\,1200~K). Aus06 report an effective temperature of 24\,000\,$\pm$\,1000~K from stellar spectra and 26\,500\,$\pm$\,500~K from the Geneva photometry. Finally, \cite{2011A&A...536L...6A} note that their spectra are consistent with $T_{\rm eff}=$ 25\,000~K. The values are therefore fairly consistent. Similarly, the reported surface gravities are compatible; $\log (g/\mbox{cm\,s$^{-2}$})=$ 3.89 \citep{2005A&A...433..659N}, 3.4\,$\pm$\,0.3 (spectra) and 3.5\,$\pm$\,0.4 (Geneva photometry, Aus06) and 3.5 \citep{2011A&A...536L...6A}. The relative metal abundance, [$m/\mbox{H}] =-$0.03 $\pm$ 0.15 \citep{2005A&A...433..659N}, indicates near-solar metallicity, which is corroborated by \citet{2011A&A...536L...6A}.

Stellar models were calculated in the same way as described by \cite{2014A&A...565A..76C}. Figure \ref{BCen20a} (left panel) shows the evolutionary tracks for stars with masses of the components of $\beta$~Cen equal to 12.0 and 10.6~M$_\odot$, calculated using OP opacities and the mass abundances equal to $X =$ 0.7374 and $Z =$ 0.0134. The value of $Z$ was taken from \cite{2009ARA&A..47..481A}, while the hydrogen to helium ratio is the same as in the line-blanketed models of \cite{2003IAUS..210P.A20C} and \cite{2012AJ....144..120M}; see \cite{2014A&A...565A..76C} for more details. We did not include core overshooting. The mixing of elements was taken into account, but only within the boundary of the convective core that was determined by the Schwarzschild criterion. The right-hand panels in Fig.~\ref{BCen20a} show ages and central hydrogen abundance, $X_{\rm c}$, as a function of model number. The same model number corresponds roughly to the same evolutionary status as can be judged from the right bottom panel of Fig.~\ref{BCen20a}, where models for both masses overlap. Models up to \#140 correspond to the core hydrogen-burning main-sequence (MS) phase, those between \#140 and \#180, to the secondary contraction, beyond \#180, to the shell-hydrogen burning post-MS phase. In Fig.~\ref{BCen20a}, we also plotted the evolutionary tracks for pseudo-rotating models in which only the spherically symmetric distortion that results from rotation is taken into account \citep{1998A&A...334..911S}.
\begin{figure*}
\centering
\includegraphics[angle=270,width=0.69\textwidth]{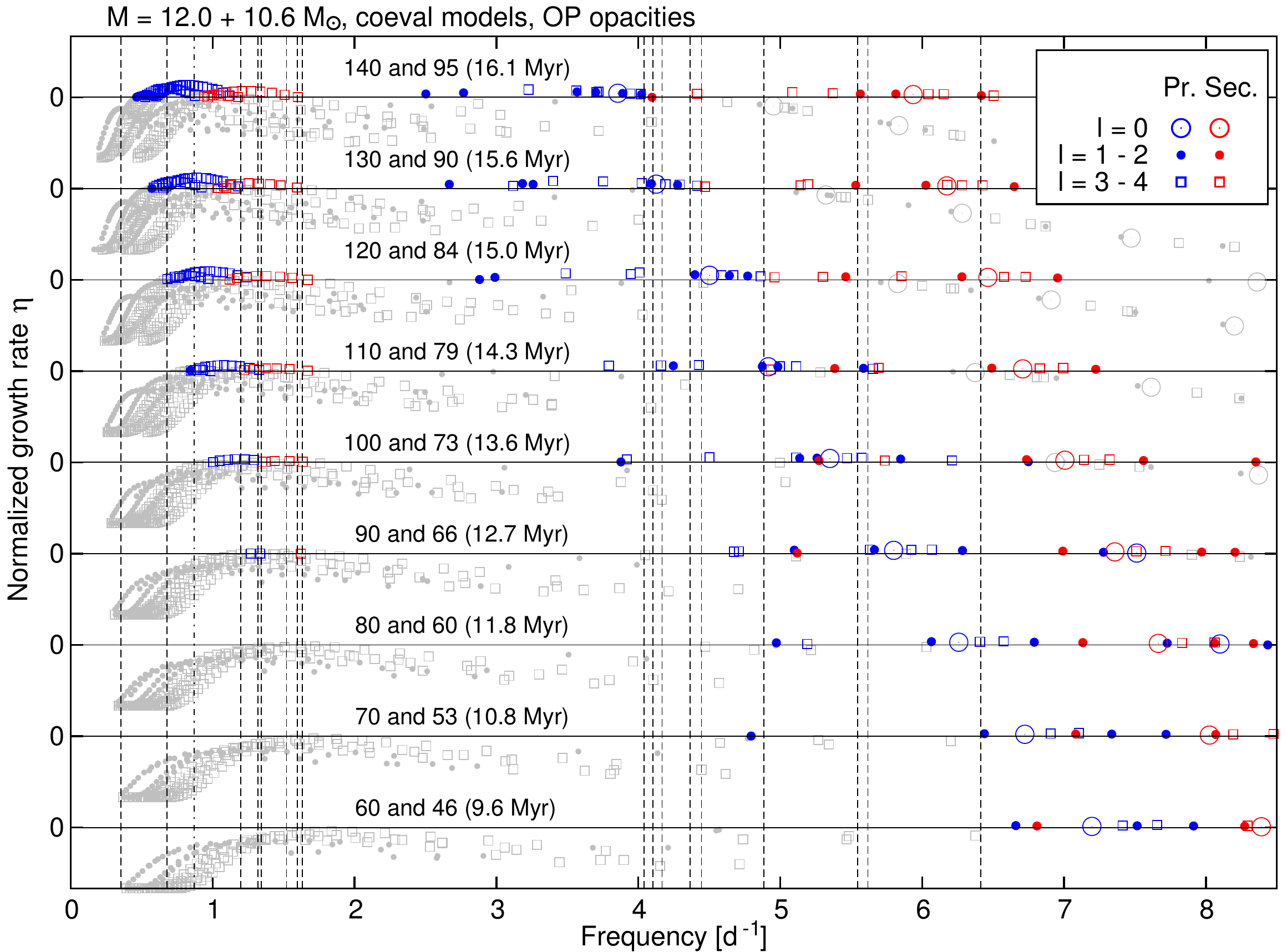}
\caption{Normalized growth rate $\eta$ \citep{1978AJ.....83.1184S} vs.~frequency for $l =$ 0\,--\,4 modes in nine pairs of coeval models with masses 12.0 and 10.6~M$_\odot$ calculated with OP opacities for $X=$ 0.7374, $Z=$0.0134 and $V_{\rm{rot}}=$ 0. The observed frequencies of $\beta$~Cen (Table \ref{freqsol}) are shown as vertical dash-dotted ($f_1$) and dashed ($f_2$ to $f_{18}$) lines. To separate the models substantially, an offset of 1.5 is introduced between the consecutive pairs. The inset explains symbols used for unstable modes with different $l$ in models of primary (Pr.) and secondary (Sec.). Stable modes ($\eta<$ 0) are shown in light gray with the same symbols as the unstable ($\eta>$ 0) modes. Pairs of models are labeled with model numbers and age. The first number corresponds to the evolutionary track of the primary, the other to that of the secondary.}
\label{eta-op}
\end{figure*}

To check which modes are unstable, we calculated seismic models using the codes of \cite{1977AcA....27..203D} for linear non-adiabatic stellar oscillations. We consider the stability of modes in the frequency range that encompasses the whole range of the observed frequencies. We focus on modes with $l\leq$ 2 but consider modes with $l$ up to 4 because as shown by \cite{1998A&A...334..911S} (see also Sect.~\ref{coupling}) rotational coupling leads to modes of mixed character and may therefore increase the visibility of the $l =$ 3 and 4 modes. We note that the two strongest $p$ modes that we detected were identified by Aus06 as having degrees $l=$ [4, 7] ($f_{3}+ \mbox{1}$~d$^{-1}$) and $l=$ [3, 5] ($f_{2} -\mbox{1}$~d$^{-1}$). We started modeling with non-rotating models. In view of the fast rotation of both components, this is a rather unrealistic evaluation. It might, however, be reasonable to look first at such a simplified example because it provides an overview of the stability of modes. The effects of rotation are discussed in the next subsections.

 In Fig.~\ref{eta-op}, we compare nine combined theoretical frequency spectra of both components with the observed spectrum of $\beta$~Cen under the assumption of coevality. The nine spectra correspond to the range of models between \#60 and \#140 for the primary and between \#46 and \#95 for the secondary. This range is in agreement with the measured effective temperatures of the bright components of $\beta$~Cen. Model \#140 lies at the terminal-age main sequence (TAMS). It is clear that none of the components alone can explain the observed frequency spectrum. In the low-frequency ($g$-mode) range only the oldest 16.1-Myr models show unstable modes in the whole observed range (0.35\,--\,1.63~d$^{-1}$), while, for models younger than $\sim$12.5~Myr, no instability in this region can be found. The unstable $g$ modes in the primary have, on average, lower frequencies than the unstable $g$ modes in the secondary. It can be also seen that $g$ modes with higher frequencies can only be explained with the $l\geq$ 4 modes that were excited in the secondary.
\begin{figure*}[!ht]
\centering
\includegraphics[width=0.75\textwidth]{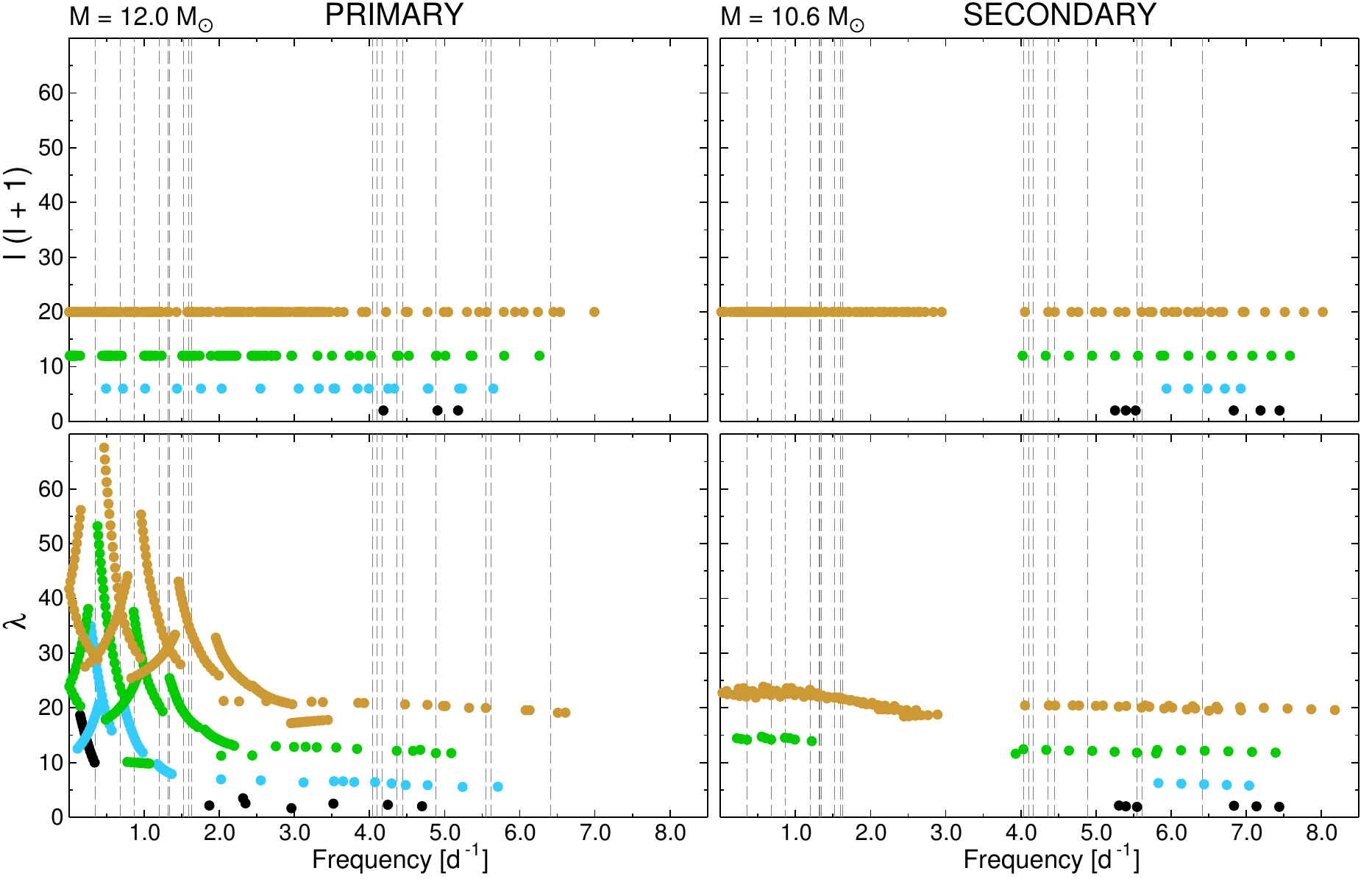}
\caption{\textit{Top panels}: Values of $l(l+\mbox{1})$ of unstable non-radial $l=$ 1\,--\,4 modes for two coeval models of the components of $\beta$~Cen, 12.0-M$_\odot$ model \#110 for the primary with $V_{\rm rot}$ = 226~km\,s$^{-1}$ and $f_{\rm rot}=$ 0.56227~d$^{-1}$ (\textit{left}) and 10.6-M$_\odot$ model \#79 for the secondary with $V_{\rm rot}$ = 94~km\,s$^{-1}$ and $f_{\rm rot}=$ 0.31746~d$^{-1}$ (\textit{right}), plotted as a function of frequency in the inertial frame of reference. \textit{Bottom panels}: Separation parameters, $\lambda$, of unstable modes for the same two models of primary (\textit{left}) and secondary (\textit{right}). The models were calculated with OP opacities for $X=$ 0.7374 and $Z=$ 0.0134. In all panels vertical dashed lines indicate the observed frequencies, the short-dashed line is the combination term $f_1$.}
\label{efrot}
\end{figure*}

\subsection{Effects of rotation}\label{fastrot}
Both components of $\beta$~Cen rotate relatively fast and therefore each attempt of seismic modeling in these stars must include the effects of rotation. First, we examined the basic properties of rotationally split components of low-degree unstable modes. We used perturbation theory that was elaborated by \cite{1998A&A...334..911S} and that includes terms that are cubic in rotation rate, $\Omega$. \cite{2006A&A...449..673S} developed a similar approach for shellular rotation. Our computing code is limited to uniform stellar rotation.

The perturbation method requires that the rotation frequency be small compared to the oscillation frequencies in the co-rotating frame of reference. This condition is accomplished at least for $p$ modes. In such a case, the observed oscillation frequency can be expressed as $\omega = \omega^0 - m \Omega + m \Omega C_L + \omega^T + \omega^D + \omega^C$, where $\omega^0$ contains the second-order correction due to spherically symmetric distortion of a pseudo-rotating star, $m \Omega$ is the Doppler effect, $m \Omega C_L$ includes the first-order correction due to rotation ($C_L$ is Ledoux constant), $\omega^T$ describes the contribution from the Coriolis second-order effect, $\omega^D$ corresponds to the non-spherically symmetric part of the Coriolis force, and $\omega^C$ arises from deviations of Coriolis motions generated by distortion. Detailed expressions for all of these terms are given by \cite{1998A&A...334..911S}.
\begin{figure*}
\centering
\includegraphics[width=0.562\textwidth]{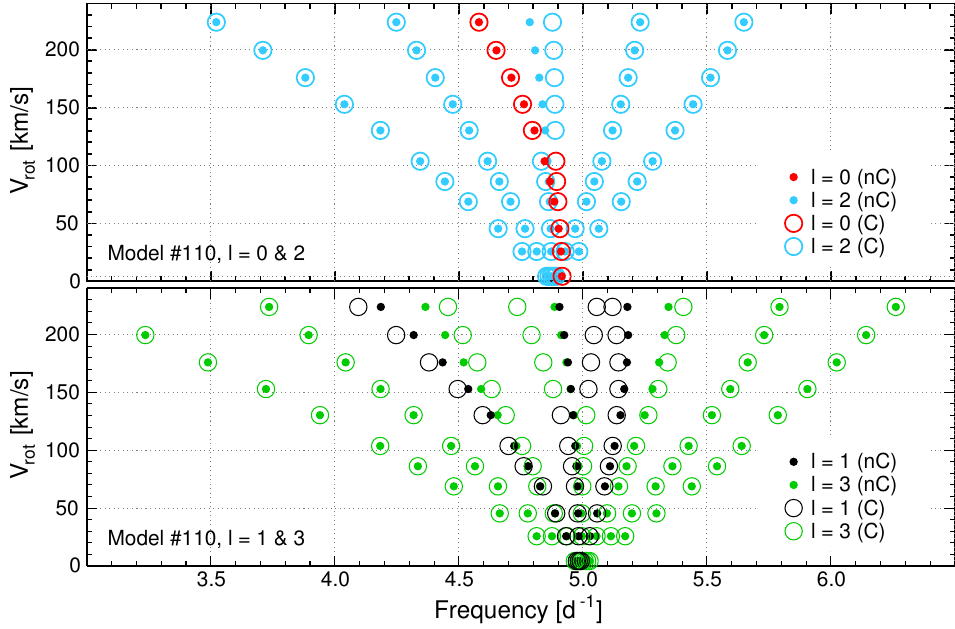}
\caption{ \textit{Top}: Rotationally coupled $l =$ 0 and 2 modes (open circles) compared with a pure rotationally split case (dots) for 12.0-M$_\odot$ model \#110. The `C' and `nC' in labels stand for the coupled and non-coupled case, respectively. \textit{Bottom}: Coupling between the $l=$ 1 and 3 modes for the same model.}
\label{coupl}
\end{figure*}

The top panels in Fig.~\ref{efrot} show unstable modes with $l=$ 1\,--\,4 for two coeval models, 12.0-M$_\odot$ model \#110 for the primary and 10.6-M$_\odot$ model \#79 for the secondary. Their frequencies are given in the inertial frame. We see that even if only modes with $l=$ 1 and 2 are considered, the range of frequencies of unstable modes in both stars encompasses the observed range. If modes up to $l=$ 4 are taken into account, the situation is even better. 

\subsection{Traditional approximation}\label{tradap}
As mentioned above, the rotational equatorial velocity of the primary component of $\beta$ Cen amounts to about 200\,--\,250 km\,s$^{-1}$. The corresponding rotational frequency for the 12.0-M$_\odot$ primary is about 0.562~d$^{-1}$ for model \#110. This frequency is comparable to the frequencies of the observed $g$ modes. In this case, the perturbation theory described in Sect.~\ref{fastrot} may be questioned. We therefore used the approach presented by \cite{2005MNRAS.360..465T}. This is the so-called traditional approximation \citep[see also][]{1998ApJ...497L.101U}. This approach allows for a separation of radial and latitudinal eigenfunctions \citep{2003MNRAS.343..125T}. An appropriate modification of non-adiabatic codes was proposed by \cite{2005MNRAS.360..465T}, \cite{2005A&A...443..557S}, and \cite{2007MNRAS.374..248D}. The high-order $g$ modes are characterized by low frequencies and their dynamics can be expected to be influenced by the Coriolis force. In the traditional approximation, the angular dependence of perturbed variables is described by the Hough functions. These are solutions of the Laplace tidal equation \citep[see, e.g.,][]{1996ApJ...460..827B,1997ApJ...491..839L}, which defines (together with boundary conditions) the eigenvalue problem on the separation parameter $\lambda$. The value of $\lambda$ changes as a function of the spin parameter $s = \mbox{2} \Omega / \omega$ for a given $m$ value, where $\omega$ is the pulsation frequency in the co-rotating frame of reference. In the limit of no rotation, the Hough function approaches the Legendre function $P_l^m$ and $\lambda = l(l + \mbox{1})$. Such eigenvalues $\lambda(s)$ correspond to rotationally modulated $g$-modes \citep{1997ApJ...491..839L}. It is therefore possible to consider branches of modes ($l, m$) for $l$ at $s = $ 0, while specification of the angular dependence additionally requires the value of $s$ \citep{2007MNRAS.374..248D}.

In this approach, the picture of instability for the same two coeval models, \#110 and \#79 (lower panels in Fig.~\ref{efrot}), is slightly different from that shown in the previous section. Now, the number of unstable modes is higher, especially in the $g$-mode domain. There is still a gap between the frequency ranges of $g$ and $p$ modes for the secondary, while there is no gap between both families of modes for the models of the primary. This is related to the higher rotation in this star. For both models shown in Fig.~\ref{efrot}, the rotational effects on the mode spectra (both $p$ and $g$) were taken into account. The frequencies of modes are also affected by rotational coupling. An example of this effect is shown in the next subsection.

\subsection{Rotational coupling}\label{coupling}
Here, we show an example of coupling of modes that was calculated using the formalism elaborated by \cite{1998A&A...334..911S}. The effect occurs when the frequencies of oscillation modes are close to each other (near-degeneracy). For B-type stars, rotation couples modes with the same $m$ but which have $l$ that is different by 0 or 2. If modes have the same $l$, modes with different radial orders $n$ can be coupled.

We considered the coupling of $l =$ 0 and 2 and, separately, $l =$ 1 and 3 modes. The effect is illustrated in Fig.~\ref{coupl} for 12.0-M$_\odot$ model \#110. As can be seen in the top panel, coupling with the radial ($l =$ 0) mode affects only the zonal ($m =$ 0) component of the $l =$ 2 mode. These two modes have mixed characters. Formally, there is a switch in frequency between the coupled radial mode and $(l,m) =$ (2, 0) mode for $V_{\rm rot}\approx$ 120\,km\,s$^{-1}$. For higher $V_{\rm rot}$ the coupling coefficients are similar for both modes, but not the same. The modes are labeled according to the higher value of the coupling coefficient of the parent mode. As indicated by \cite{1998A&A...334..911S}, the resonance pushes the frequencies apart if they approach each other too closely. Consequently, observed frequencies may differ significantly when no coupling is considered.

A similar effect occurs for $l =$ 1 and 3 modes, shown in the bottom panel of Fig.\,\ref{coupl}. Now three pairs of modes ($m = -$1, 0, and $+$1) are coupled. The coupling markedly influences their frequencies for $V_{\rm rot}\approx$ 200~km\,s$^{-1}$, but is already non-negligible for $V_{\rm rot}=$ 100~km\,s$^{-1}$. If more than two modes are close to one another, the behaviour may be much more complicated than reported here.

\subsection{Magnetic field of the secondary}
As discussed in Sect.~\ref{aaab}, \citet{2011A&A...536L...6A} detected the presence of a magnetic field in the Ab component of $\beta$~Cen. Few detailed asteroseismic studies have been performed on magnetic $\beta$~Cep stars \citep{2000ApJ...531L.143S,2012MNRAS.427..483B}. For MS stars, the theoretical work on pulsation-magnetic field interaction has generally been focused on less massive roAp stars \citep[e.g.,][]{1995PASJ...47..219T,2004MNRAS.350..485S} so, in principle, it is currently unknown how the presence of a magnetic field will modify the observed frequencies in hot pulsators. This therefore opens up an interesting discussion on the role of the interaction of magnetic fields with pulsations in early-type stars. Pulsations and magnetic fields might be also related to X-ray variability that has been discovered in some stars \citep{2014NatCo...5E4024O,2014ApJS..215...10N}. However, only weak evidence for X-ray variability was found in $\beta$~Cen \citep{1997ApJ...487..867C,2005A&A...437..599R,2015A&A...577A..32O}. In general, however, the presence of a magnetic field can be regarded as another complication in the interpretation of the pulsation spectrum of $\beta$~Cen.

\section{Discussion and conclusions}\label{discon}
The B-C data of the massive triple $\beta$~Centauri analysed in the present paper clearly shows the potential of the B-C mission for in-depth studies of massive pulsating variables. With 17 detected modes (both $p$ and $g$), this binary star becomes one of a few known hybrid $\beta$~Cep/SPB stars with such a rich frequency spectrum. On the one hand, the binarity of $\beta$~Cen provides a very precise determination of the masses of the components, which is the most important parameter of a model. On the other hand, however, it complicates seismic modeling because the modes need to be not only identified but also assigned to one of the components. Attempting this kind of assignment, using the rates of frequency change (Sect.~\ref{aaorab}), was not conclusive, but there is a chance that with more data from the B-C this may become possible for the $p$ modes. 

The possibility of mode identification (MID) using two-band observations was indicated many times as one of the main advantages of the B-C mission, compared to other satellites doing photometry from space. While, in general, this is true, the blue-to-red amplitude ratios for $\beta$~Cen are likely to be affected by non-linearity and should therefore not be used for this kind of diagnostic. All we can say is that amplitude ratios $A_B/A_R$ of $f_2$ and $f_3$ modes are significantly different from those of $f_5$, $f_6$ and $f_{10}$, so that the modes probably have different $l$. The instrumental effects do not affect phase differences, which are given in Table \ref{freqsol}. Unfortunately, owing to the small amplitudes, the uncertainties of phases are quite large and, therefore, phase differences are of limited use as well.

The main complication in seismic modeling is the relatively fast rotation of both components. In view of the effects shown in Sects \ref{fastrot} and \ref{tradap}, it is clear that MID cannot be achieved with the frequencies themselves. Fortunately, there is a large collection of high-resolution spectra in the public databases which should allow for the detection and identification of some modes not only in terms of $l$, but also $m$, which is crucial in seismic modeling of rapidly rotating stars. This kind of analysis, which should also include spectral line modeling similar to that of \cite{1997MNRAS.284..839T} is, however, beyond the scope of this study and is postponed to a future paper. The rapid rotation of the $\beta$~Cen components seems to be the most severe complication for seismic modeling, but on the other hand we need more rapidly rotating stars to confront predictions of the current theory of stellar pulsation. The study of stars similar to $\beta$~Cen can shed some light on our understanding of pulsations in rapid rotators, even such fast rotators as Be stars. The B-C mission seems to be perfectly suited for this kind of study and there are at least several other rapidly rotating $\beta$~Cep stars ($\delta$~Lup, $\epsilon$~Cen, $\epsilon$~Per) that have already been observed by the B-C.

As shown in Sect.~\ref{parmod}, non-rotating models have problems in explaining the hybridity of $\beta$~Cen. The problem is general as indicated in the Introduction: to make $g$ modes unstable in some slowly rotating $\beta$~Cen/SPB hybrids, a change of opacity was postulated. Not concluding on the need for such a change, we would like to point out that we get unstable modes in the whole range in which they are observed in $\beta$~Cen, if effects of rotation are taken into account (Sect.~\ref{tradap}, Fig.~\ref{efrot}). In addition, there is no need to increase the metallicity or change the chemical composition.

\begin{appendix}
\section{A correction of raw B-C data for instrumental effects}\label{prep}
As we mentioned in Sect.~\ref{br-phot}, raw B-C photometry requires several preparatory steps before it can be subjected to time-series analysis. These steps are as follows:\footnote{Check the BRITE Wiki page (http://brite.craq-astro.ca) for updates of the procedure.} (i) rejection of outliers, (ii) rejection of corrupted orbits, (iii) decorrelations, and (iv) merging the light curves from different BRITE satellites. We describe them in detail in this appendix using $\beta$~Cen data as an example.

\begin{figure}[!ht]
\centering
\includegraphics[width=0.45\textwidth]{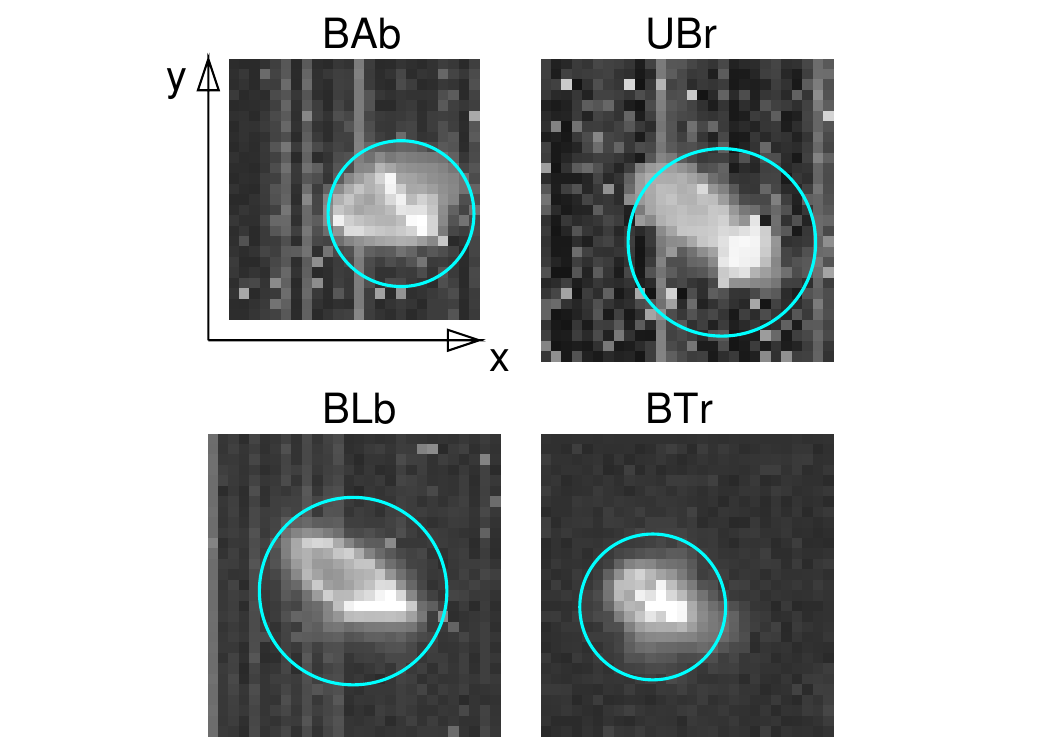}
\caption{\small Sample raster frames of $\beta$~Cen from the four BRITEs: BAb (top left), UBr (top right), BLb (bottom left), and BTr (bottom right). The optimal apertures (shown as circles) are shown. Their radii are equal to 7 pixels for BAb and BTr and 9 pixels for BLb and UBr. The gray scale in not linear. Note the smaller raster size (24 $\times$ 24 pixels) for BAb observations compared to the rasters from other satellites (28 $\times$ 28 pixels).}
\label{frames}
\end{figure}

The B-C photometry is provided in the form of text files with the results of aperture photometry corrected only for intra-pixel variability. The files contain several columns, namely: heliocentric Julian day; the total signal measured through the optimal circular aperture, $F_{\rm tot}$; two coordinates of the center of gravity of a stellar image in a raster, $x_{\rm cen}$ and $y_{\rm cen}$; chip temperature, $T$; and -- in the case of $\beta$~Cen -- two other parameters which measure the degree of smearing and which will be explained below. The aperture center is the center of gravity of a thresholded stellar profile which -- because of defocusing -- can be very complicated and varies for different satellites, see example profiles in Fig.~\ref{frames}. The optimal aperture radius, $r_{\rm opt}$, is derived during a procedure in which the median absolute deviation (MAD), which is obtained for aperture radii in the range between 3 and 10 pixels, is minimised. Another parameter optimised during the reduction step is the threshold for the identification of bad pixels in the mean dark frame, $S_{\rm th}$. In the course of the reduction, the images are also noise-reduced by a column offset removal and hot pixel replacement. The filtering scheme utilises the fact that a star changes its position in its raster owing to the imperfect tracking of the satellite. Therefore, bad pixels may be retrieved from other images in a series (see Popowicz et al., in preparation, for more details). The final photometry is that which gives the lowest MAD from all time series that are obtained with different $r_{\rm opt}$ and $S_{\rm th}$. Sample images of $\beta$~Cen from the four BRITEs together with optimal apertures are shown in Fig.~\ref{frames}. 

\begin{figure}[!ht]
\centering
\includegraphics[width=0.45\textwidth]{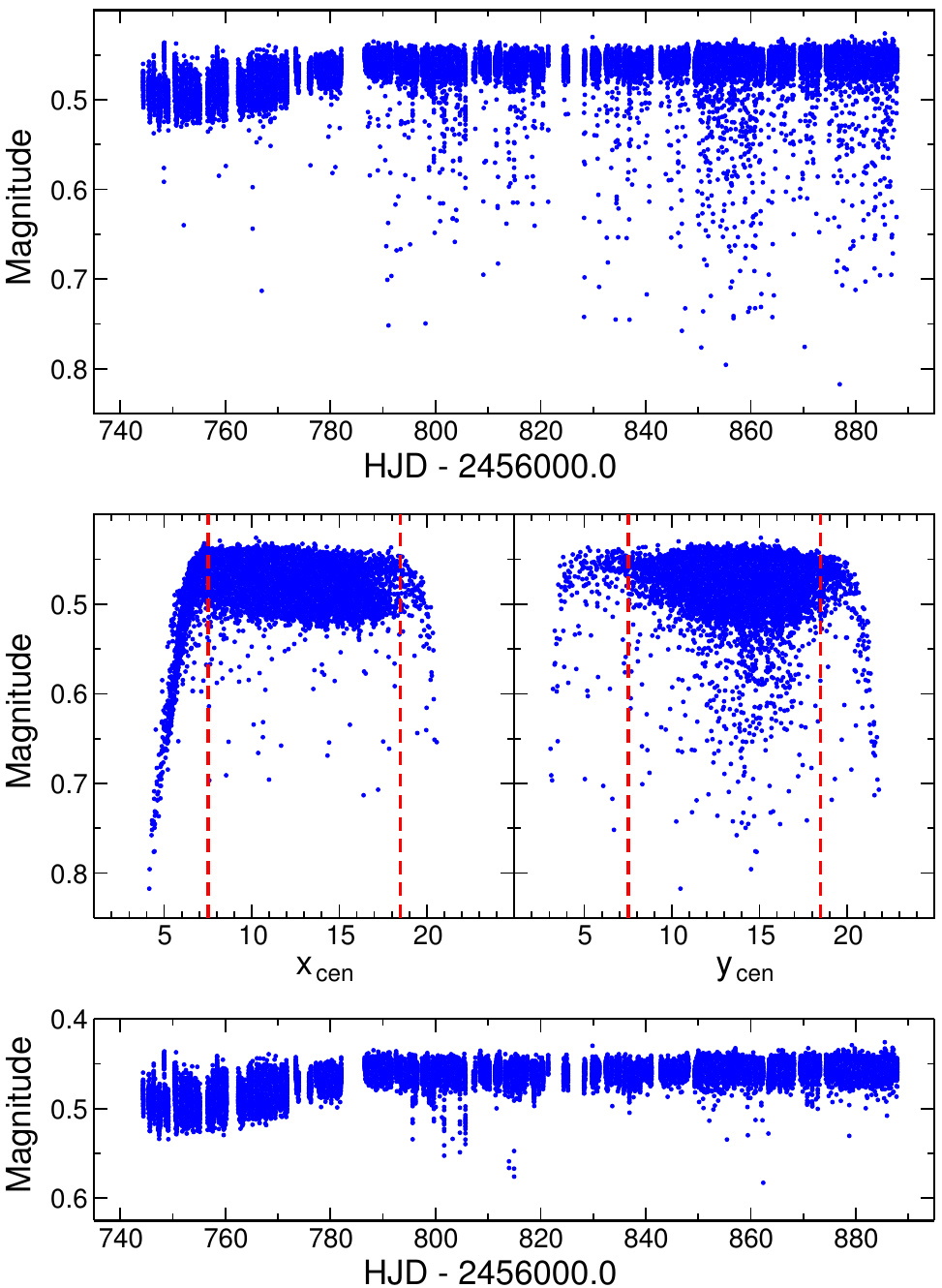}
\caption{\small {\it Top}: The raw BAb light curve of $\beta$~Cen. {\it Middle}: Raw magnitudes plotted as a function of $x_{\rm cen}$ ({\it left}) and $y_{\rm cen}$ ({\it right}). Vertical dashed lines encompass the ranges of $x_{\rm cen}$ and $y_{\rm cen}$ in which the raster fully includes the optimal aperture. {\it Bottom}: The BAb light curve of $\beta$~Cen after removing data points that come from frames with aperture extending beyond the raster edge.}
\label{raw-lc}
\end{figure}

\subsection*{Outlier removal}\label{rem-out}
At the beginning of the preparatory procedure, the raw fluxes are converted to magnitudes according to the formula 
$$m = -\mbox{2.5}\log F_{\rm tot} + \mbox{14.5},$$ 
where the constant 14.5 has been chosen arbitrarily. The raw BRITE photometry from one of the satellites (BAb) is shown in Fig.~\ref{raw-lc}. Two effects can be seen in this light curve: an instrumental change of brightness during the first $\sim$50 days (this is explained below) and a large number of outliers. Most of the outliers deviate in one direction, towards a lower flux. Their appearance can easily be explained when raw magnitudes are plotted as a function of $x_{\rm cen}$ and $y_{\rm cen}$ (middle panels of Fig.~\ref{raw-lc}). Clearly the outliers come from the sudden drops in magnitude for the lowest and the highest $x_{\rm cen}$ and $y_{\rm cen}$. These drops appear when the optimal aperture extends beyond the edge of a raster and therefore some amount of flux is lost. These data points appear in the final photometry because the corresponding images are still useful for hot pixel replacement and in the procedure for deriving the optimal apertures that is performed as a part of the reduction process. Normally, one should remove all data points that were obtained from images in which the observed raster does not encompass the whole aperture (points outside the two vertical dashed lines in the middle panels of Fig.~\ref{raw-lc}). Sometimes, however, if the aperture does not extend much beyond the raster edge, the amount of flux which is lost can be negligible. This depends on the shape of the stellar profile and the direction in which a star is shifted, for example when the stellar profile does not fill the lower part of the aperture in the BAb frame (Fig.\,\ref{frames}). Therefore, shifting the aperture downwards (lower $y$) would not lead to a perceptible loss of flux even if the aperture extends slightly beyond the raster. This is exactly what can be seen in the magnitude vs.~$y_{\rm cen}$ plot (middle right panel of Fig.~\ref{raw-lc}): for low $y_{\rm cen}$, the drop of magnitude starts only when the aperture extends several pixels beyond the raster edge. The conclusion here is that one may consider keeping the data points that were obtained from frames in which the incomplete aperture is covered, especially because the effects of a small flux loss can be compensated for in the procedure of decorrelation (see below).

Having removed the data points obtained from frames in which the aperture extends beyond the raster, we are left with the light curve which is shown in the bottom panel of Fig.~\ref{raw-lc}. Some outliers are still left but they have a different origin. The main reason is an imperfect tracking which leads to the motion of a star during an exposure and consequently smears the stellar image. Since a constant circular aperture is used, the smearing causes a loss of a part of the flux. In general, BLb and BTr data show a much lower number of outliers than BAb and UBr because they have better trackers \citep{2014PASP..126..573W}.

For the $\beta$~Cen data, the average number of data points per orbit is 31, 27, 43 and 87 for BAb, BLb, UBr, and BTr, respectively. This allows one to use orbit samples for a statistically justified and efficient outlier removal. For this purpose, we implemented the Generalized Extreme Studentized Deviate (GESD) algorithm \citep{1983Techn..25..165R} but, in principle, any method to reject outliers in a robust way can be applied (e.g., sigma clipping). The GESD is parametrised by a single parameter $\alpha$. The larger $\alpha$, the more outliers are detected, up to the limit of maximum percentage of outliers allowed that needs to be set in the algorithm. The decision whether to apply outlier removal, and how many times to perform this procedure, is somewhat arbitrary. In general, we applied such filtering after several steps of decorrelation, in an attempt to preserve as many points as possible.

In view of the further steps, especially decorrelations, an important question arises: should intrinsic variability be removed prior to the outlier removal and decorrelations, or not? If the intrinsic variability has very small amplitudes, it is not necessary to remove the variability before the decorrelations. For stars that show intrinsic variability with amplitudes comparable to the scatter of points or higher, this would be a mandatory step, however. In this case, each correction to the data must be followed by an appropriate fitting and all the corrections should be derived using residuals from the fit. It is enough to use a solution which includes only the strongest periodicities. For example, for $\beta$~Cen our fit included only the two strongest periodic terms, $f_1$, and $f_2$.

\subsection*{Removal of the corrupted orbits}\label{orbits}
\begin{figure}[!ht]
\centering
\includegraphics[width=0.45\textwidth]{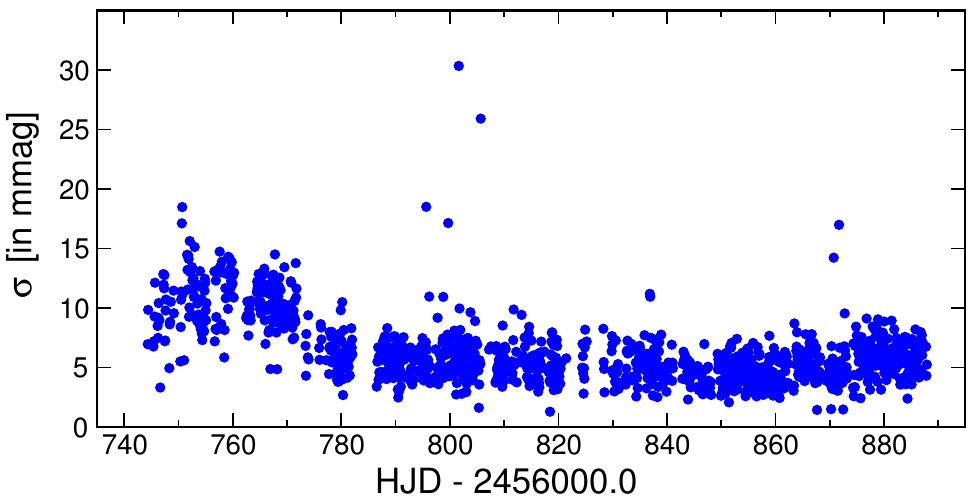}
\caption{\small Standard deviations $\sigma$ in orbit samples for the BAb data of $\beta$~Cen before decorrelations. Note the presence of orbits with much higher scatter.}
\label{orb}
\end{figure}

For some reason, usually because of much worse tracking or stray light, the data in some orbit samples (typically covering 10\,--\,20 minutes) showed much larger scatter than in the others (Fig.~\ref{orb}). In addition to the removal of outliers, one should therefore consider removing the whole orbit samples of data that exhibit large scatter. This removal can be done at any step of the preparatory procedure and, if necessary, may be repeated several times. This is because decorrelations may change (decrease) the scatter in different orbits to a different extent. For $\beta$~Cen, the removal of the corrupted orbits was done three times for BAb and UBr data and once for the BLb data. No removal was needed for the BTr data.

\subsection*{Decorrelations}\label{decor}
\begin{figure}[!ht]
\centering
\includegraphics[width=0.45\textwidth]{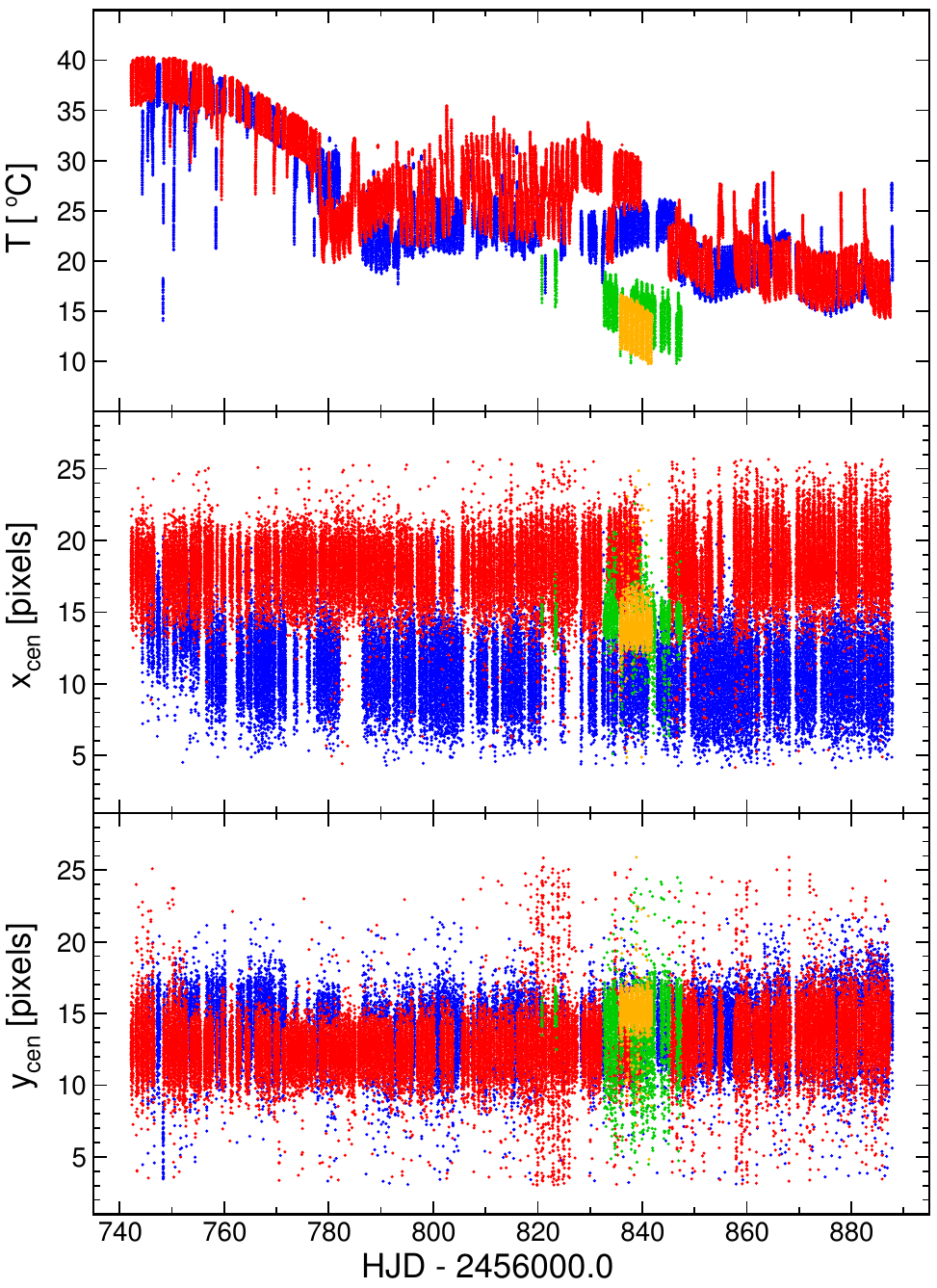}
\caption{\small Changes of $T$ (top), $x_{\rm cen}$ (middle) and $y_{\rm cen}$ (bottom) for $\beta$~Cen observations by BAb (blue), UBr (red), BLb (green), and BTr (orange).}
\label{corpar}
\end{figure}

The raw BRITE magnitudes correlate, sometimes strongly, with several parameters, namely: the position of the stellar centroid defined by $x_{\rm cen}$ and $y_{\rm cen}$, chip temperature, orbital phase and the degree of smearing. The correlations therefore form the most significant instrumental effect in BRITE photometry. Why do they occur? Let us start with the temperature of a CCD chip. In most ground-based systems, CCDs are cooled and their temperature is stabilised. For lack of space, there is no such cooling installed onboard BRITEs, so that the chip temperature reflects the thermal response of the whole satellite for illumination by the Sun and internal heating. The variable temperature of a BRITE CCD chip leads to the occurrence of a temperature effect. The change of chip temperature for the Centaurus field data is depicted in Fig.~\ref{corpar}, top panel. The character of these changes depends mainly on the satellite's orbit and orientation. Despite the long-term changes of temperature, there is also a change of temperature within a single orbit. Its range corresponds to the vertical extent of a strip and the temperature usually increases. By comparing temperature changes for BAb in Fig.~\ref{corpar} and raw magnitudes in Fig.~\ref{raw-lc}, one can see that the magnitude drop at the beginning of BAb observations likely comes from the temperature effect.

Although the images cover a number of pixels, the positions of the stellar centroid vary in a significant range. In this situation, the lack of flat-fielding (there are no large enough uniform sources of illumination available and downloading the whole chip takes more than a day) leads to a strong dependence of the measured flux on the centroid position, $x_{\rm cen}$ and $y_{\rm cen}$. Even though the intra-pixel sensitivity correction is applied in the main photometric routine, some residual dependencies remain (seen as a one-pixel long wave). It is also obvious that the better the stability of a satellite, the more accurate the photometric outcome. This is the main reason why photometry from better stabilized BLb and BTr usually has smaller scatter than from BAb and UBr. The other factor which makes the BLb and BTr photometry less scattered is the lower total radiation dose, which they were exposed to and, consequently, a smaller number of hot pixels (Pablo et al., in preparation).

In addition to the dependence on $T$, $x_{\rm cen}$, and $y_{\rm cen}$, one can expect to see instrumental effects related to the orbital phase, $\phi$, as can be observed in data collected by many satellites working in low-Earth orbits. It seems that for BRITEs this is a rather complicated dependence causing the occurrence of several peaks close to the orbital frequency and its harmonics. The orbital frequencies of BRITEs differ a little and can slightly change with time. For the Centaurus field observations they amount to 14.340, 14.339, 14.46, and 14.65~d$^{-1}$ for BAb, UBr, BLb, and BTr, respectively. Decorrelation with $\phi$ decreases the amplitude of the orbital frequency and its harmonics, but usually does not remove all orbit-related instrumental signals.

For most stars observed by the BRITEs, decorrelations with the four parameters which are either provided in the data file ($T$, $x_{\rm cen}$, $y_{\rm cen}$) or can be easily calculated ($\phi$) were sufficient to remove practically all instrumental effects. The star, $\beta$~Cen, however, is very bright ($V=$~0.6~mag). After the decorrelations with the four parameters some brightenings occurred in the light curve (especially in the UBr data), which were apparently of instrumental origin. A careful inspection of the associated frames revealed that these were smeared images. As we already mentioned, smearing occurs as a result of imperfect tracking and should therefore cause a loss of flux measured in a constant aperture. However, we observed the opposite effect: the smeared images resulted in a higher total flux. This can be explained only by the non-linearity at a higher signal level. This is a result of the anti-blooming protection which is used in BRITE CCD chips (Pablo et al., in preparation). To correct for this additional effect, several parameters measuring the amount of smearing were proposed. Ultimately, two such parameters were chosen and the data for $\beta$~Cen were re-reduced to calculate them for all the frames. The two parameters, let us call them $A$ and $B$, are as follows: 
\begin{equation}
A=\sum_{i=1}^N \left({S_i\over S_N}\right)^2,\quad S_N = \sum_{i=1}^{N} S_i,
\end{equation}
where $S_N$ is the sum of signals from all pixels in the aperture, $S_i$, the signal in the $i$th pixel, and summing is over $N$ pixels falling into the aperture. In other words, $A$ is the sum of squared normalized intensities. This quality metric is based on the image energy, which is largest when an image is sharp and flux is spread over a small CCD area. The other parameter, $B$, is defined as
\begin{equation}
B={\mbox{1}\over{\mbox{4}S_M^2}}\sum_{k=1}^4 \sum_{i=1}^{L} S_i S_{i,k},\quad S_M = \sum_{i=1}^{M} S_i,
\end{equation}
where $S_M$ is the sum of signals from all $M$ pixels in a raster, $S_{i,k}$ is the signal in the $i$th pixel in the image that has been shifted by one pixel in one of the four ($k$) directions: up, down, left and right. The summing is over $L$ pixels, which the image has in common with its shifted copy. This type of smearing assessment is based on the correlation of the image with its shifted copies. If smearing appears and the object is blurred, the correlation increases. Consequently, the sharp, high-fidelity, point-spread function produces distinctly lower $B$. The two parameters, $A$ and $B$, are correlated, whereby decorrelation with one of them also considerably decreases the correlation with the other. 
\begin{figure}[!ht]
\centering
\includegraphics[width=0.42\textwidth]{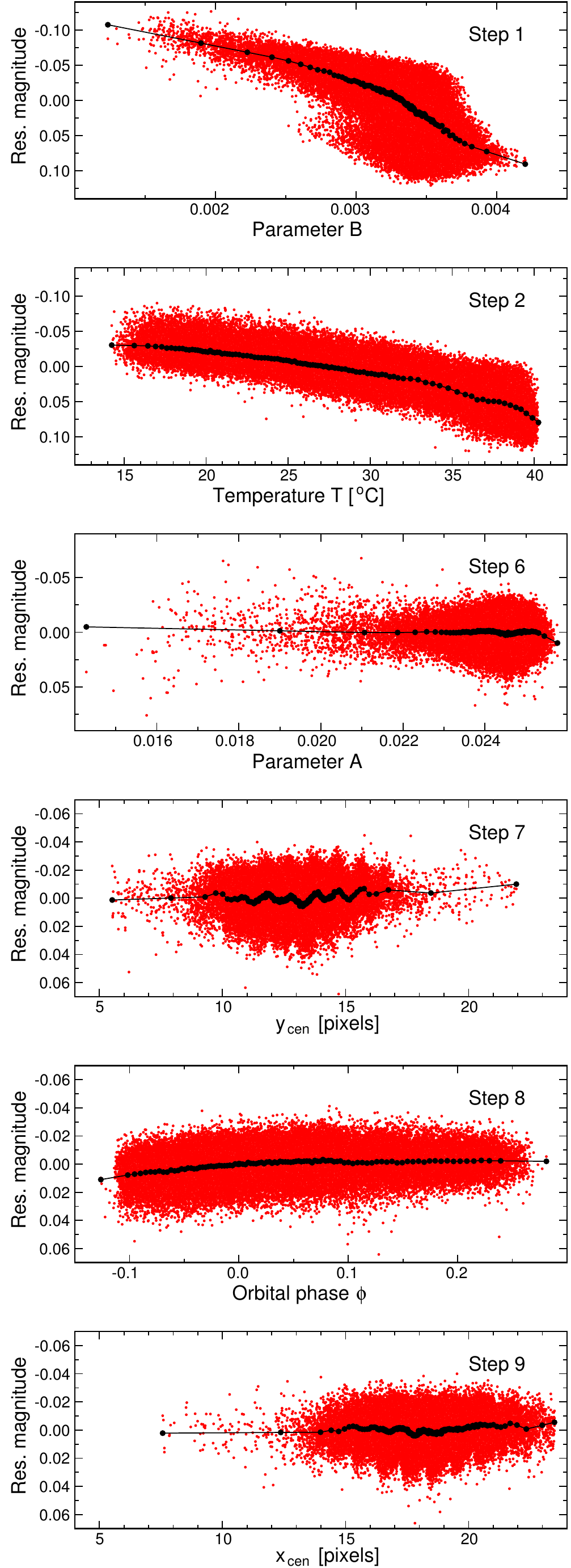}
\caption{\small Six selected steps of decorrelation for the UBr data of $\beta$~Cen. The panels show residual magnitude as a function of a given parameter (going from top to bottom, $B$, $T$, $A$, $y_{\rm cen}$, $\phi$, and $x_{\rm cen}$). The black dots are the anchor points.}
\label{decor-BC}
\end{figure}

The final decorrelation procedure for $\beta$~Cen data therefore included six parameters. It would be best to do the multi-dimensional decorrelation in one step. However, as we show below, the correlations cannot be described by a simple function. In addition, the necessity of applying outlier and/or orbit removal makes the whole procedure interactive. Therefore, decorrelations have been performed sequentially. At each step of the decorrelation, correlations with all parameters were checked by calculating averages in intervals of a given parameter. These averages were next used as the anchor points for Akima interpolation \citep{1991ACMT...17..341A,1991ACMT...17..367A} and the ranges of correlations were derived. The parameter which showed the strongest correlation was used in the next step of the decorrelation. Sometimes if correlations were strong, the procedure required multiple decorrelations with the same parameter. The process was stopped when there was no more significant correlation with any of the six parameters. 

Figure \ref{decor-BC} shows six steps of decorrelation for the UBr data of $\beta$~Cen. The whole procedure for this data set included the following sequence of 17 decorrelation steps: $B-T-B-T-B-A-y_{\rm cen}-\phi-x_{\rm cen}-T-A-y_{\rm cen}-\phi-B-A-x_{\rm cen}-T$. In general, the required number of decorrelation steps strongly depends on the data set, star, and satellite. For the $\beta$~Cen data, the whole decorrelation required 13, 17, 11, and 12 steps for BAb, UBr, BLb, and BTr, respectively. For all data sets, decorrelation with $B$ was one of the most significant, indicating that for a very bright star the non-linearity effect is important. Fainter stars observed by the BRITEs did not show this effect.

The importance of decorrelations can be judged from Fig.~\ref{decor-BC}, but it can be also evaluated from the comparison of scatter before and after this step. When standard deviations of the complete data sets from a given satellite are compared, then the result is the following (for $\beta$ Cen data): decorrelation reduces the standard deviation by a factor of 2.41, 4.36, 2.00, and 1.77 for BAb, UBr, BLb, and BTr, respectively. The correlations are strongest for the UBr data, but are significant for all satellites. A comparison of Fig.~\ref{B-data} with Fig.~\ref{raw-lc} clearly shows that decorrelations account for the instrumental long-term drift. For the BAb data of $\beta$~Cen, it is the temperature effect which is responsible for the instrumental brightening at the beginning of the run. If median standard deviations per orbit before and after decorrelation are compared, then the reduction factors amount to 1.15, 1.34, 1.47, and 1.75 for BAb, UBr, BLb, and BTr, respectively. A general conclusion is therefore the following: Decorrelation is a crucial step in the preparatory procedure and can significantly improve the photometry. This is especially important when -- as in our case -- sub-mmag signals are searched for.

Finally, we would like to make an important note on intrinsic variability: if it is periodic, then it may be easily subtracted. If not, it may be impossible to unambiguously separate intrinsic variability and instrumental effects. In particular, separation of the long-term variability and correlation with $T$ might be difficult.

\subsection*{Data merging}
The last step in the preparatory procedure is data merging. As in Sect.~\ref{tser}, we may want to combine data from different satellites that are observing with the same (blue or red) filter or even combine data that is obtained with blue and red filters. An obvious step before doing this is to subtract the mean magnitudes, which are easy to get from the fit if a star shows (multi)periodic intrinsic variability. If long-term variability is present, this is challenging and depends on whether the data overlap or not. In the latter case, unambiguous merging cannot be done. Prior to merging, one may also want to remove the remaining instrumental variability which was not accounted for by decorrelations. This kind of variability manifests itself by peaks at frequency of 1 or 2~d$^{-1}$ or close to the orbital frequency or its multiples. It is therefore reasonable to perform time-series analysis separately for each data set, remove instrumental variability, if present, and then merge data from different sets altogether.
\end{appendix}

\begin{acknowledgements} This work used the SIMBAD and Vizier services operated by Centre des Donn\'ees astronomiques de Strasbourg (France), biblio\-graphic references from the Astrophysics Data System maintained by SAO/NASA, and the Washington Double Star Catalog maintained at USNO. The work was supported by the Polish NCN grants no.~2011/03/B/ST9/02667, 2013/11/N/ST6/03051, and 2011/01/B/ST9/05448, and the BRITE PMN grant 2011/01/M/ST9/05914. A.F.J.\,Moffat was supported by the NSERC, FRQNT, and CSA grants. S.M.\,Rucinski, J.M.\,Matthews, and G.A.\,Wade acknowledge research grants from Natural Sciences and Engineering Research Council (NSERC) of Canada. G.\,Whittaker, H.\,Pablo, and T.\,Ramiaramanantsoa acknowledge support from the Canadian Space Agency grant FAST. R.\,Kuschnig and W.W.\,Weiss acknowledge support from an ASAP9 grant provided by FFG. K.\,Zwintz acknowledges support by the Austrian Fonds zur F\"orderung der wissenschaftlichen Forschung (FWF, project V431-NBL). We thank the anonymous referee for useful comments.
\end{acknowledgements}

\end{document}